\documentclass[a4paper,10pt]{article}
\usepackage{amsmath}
\usepackage{amsfonts}
\usepackage{amssymb}
\usepackage[dvips]{graphicx}

\newcommand{\cyl}[1]{$\Delta^*_{\varepsilon (#1)}$}
\newcommand{\boup}{S^{(+)}_{\varepsilon (p)}}
\newcommand{\boum}{S^{(-)}_{\varepsilon (p)}}

\newcommand{\fa}{\mathfrak{a}}
\newcommand{\ofa}{\overline{\mathfrak{a}}}

\newcommand{\ie}{\emph{i.e.}}
\newcommand{\q}{e^{\frac{2\pi}{2\pi -\varepsilon(p)}}}
\newcommand{\Rangle}{\rangle\negthinspace\rangle}
\newcommand{\ordprod}[1]{:\negmedspace{#1}\negmedspace:}

\newcommand{\bgz}{\bar{\zeta}}

\newtheorem{theorem}{Theorem}
\newtheorem{proposition}[theorem]{Proposition}

\begin{document}

\begin{titlepage}
  \begin{flushright}
    FNT/T 2007/03\\
QMUL-PH-07-12
  \end{flushright}
\begin{center}
{\Large \textbf{Boundary Conformal Field Theory\\ and Ribbon Graphs:}}

\bigskip

{\Large \textbf{a tool for open/closed string dualities}}

\vspace{0.5 cm}

{\large \textsl{Mauro Carfora}$^{a,}%
$\footnote{{email: mauro.carfora@pv.infn.it}},}
{\large \textsl{Claudio Dappiaggi}$^{a}$%
\footnote{{email: claudio.dappiaggi@pv.infn.it}},} 
{\large \textsl{Valeria L. Gili}$^{a,b,}$%
\footnote{{email: V.Gili@qmul.ac.uk}}.}

\vspace{0.5 cm}

\vspace{0.1cm}

$^{a}$~Dipartimento di Fisica Nucleare e Teorica,\\
Universit\`{a} degli Studi di Pavia, \\[0pt]
and\\[0pt]
Istituto Nazionale di Fisica Nucleare, Sezione di Pavia, \\[0pt]
via A. Bassi 6, I-27100 Pavia, Italy

\vspace{0.2cm}

$^{b}$~Centre for Research in String Theory\\
Department of Physics\\
Queen Mary, University of London\\
Mile End Road\\
London E1 4NS, UK

\vspace{0.1cm}

\bigskip
\end{center}

\begin{abstract}
  We construct and fully
  characterize a scalar boundary conformal field theory on a
  triangulated Riemann surface. The results are analyzed from a string
  theory perspective as tools to deal with open/closed string
  dualities.
\end{abstract}

\vfill

\textbf{PACS}: 04.60.Nc, 11.25.Hf.
\textbf{AMS Subj. Class.}: 83C27

\textbf{Keywords}: Boundary conformal field theory,
Discretized models of gravity, String dualities

\end{titlepage}

\section{Introduction and motivations}

The dynamical description of the open--to--closed worldsheet
transition is a delicate issue in the analysis of open/closed string
dualities. Simplicial techniques play, in this setting, an important
role providing a deep, unexpected connection between Riemann moduli
space, conformal field theory, and the study of the gauge/gravity
correspondence
\cite{Gai-Ras,Gopakumar:2003ns,Gopakumar:2004qb,Gopakumar:2005fx,Aharony:2006th,David:2006qc,Aharony:2007fs}.
The kinematical rationale motivating such a role is provided by the
ribbon graph realization of gauge theory diagrams
\cite{Gopakumar:2005fx}, by the Schwinger parametrization of the
polytopal cells of the graph, and by Strebel's theorem
\cite{Strebel,mulase}, connecting the combinatorics of decorated
ribbon graphs to the conformal geometry of the worldsheet. Even if
this suggests that we are disclosing some deep discrete structure
underlying string dualities, we must stress that the dynamical aspects
of the connection between combinatorial structures and dualities is
not so obvious. Here, Boundary Conformal Field Theory (BCFT) is called
into play in a rather sophisticated way: the ribbon graph, and more
generally the underlying (metrically) triangulated surface, becomes
the combinatorial pattern along which the quantum matter fields,
described by cell--wise independent BCFTs, interact. One expects that
this interaction generates a BCFT living on an open Riemann surface
with gauge decorated boundaries living on D--branes which act as
sources of the gauge fields. The actual realization of such a BCFT is
notoriously difficult to carry out explicitly, and a careful analysis
of its construction is the main motivation underlying the analysis
presented in this paper.

\noindent Whereas our set up will be necessarily rather elementary on
the CFT side, since we consider bosonic matter fields, it will be
geometrically quite general in the sense that we consider metric
ribbon graphs which are dual to triangulated surfaces with curvature
defects. The reason of this generality is that curvature defects
provide a natural order parameter which allows to map closed
$N$--pointed Riemann surfaces into open Riemann surfaces with $N$
boundary components (of definite lengths). Such a mapping has been
described in details in \cite{Carfora1,Carfora2,Carfora3} where metric
triangulations with variable connectivity and variable edge lengths,
the Random Regge Triangulations (RRT), have been considered. It also
has an equivalent description in terms of hyperbolic geometry
\cite{Carfora:2006nj}, which is naturally activated if one consider
matter in the form of twistorial fields. For simplicity, we limit here
our analysis to the more elementary RRT case. In such a setting, a
basic step is to exploit the fact that a RRT with curvature defects
can be naturally uniformized on an open Riemann surface, with finite
cylindrical ends whose moduli are provided by the defect
\cite{Carfora1}. These finite cylindrical ends are glued together
along the pattern defined by the ribbon graph baricentrically dual to
the parent triangulation. One can naturally interpret each cylindrical
end as an open string connected at one boundary to the ribbon graph
associated to the discretized worldsheet, while the other boundary
lies on a D-brane acting as a source for gauge fields. The main topic
we address here concerns the coherent description and definition of a
BCFT - (\ie{} an open string theory in a worldsheet meaning) on such a
background.  In particular, at a fixed genus $g$ and at a fixed number
of vertexes $N_0$ in the underlying simplicial complex, we first
quantize a $D$-dimensional BCFT on single cylindrical end.  Then, we
will show how the resulting theories on different cylinders can be
glued together along the intersection pattern defined by the ribbon
graph associated to the given RRT (for some preliminary results in
this direction see \cite{Carfora:2004fd}).  This latter aspect, which
is the main result of this paper, calls into play BCFT in a non
trivial way and it is based upon a careful use of the automorphisms of
the chiral algebra naturally associated with the conformal theory. The
generality of the overall construction allows us to take in account
also open string gauge degrees of freedom, since the outer boundary of
the cylindrical ends can lay on a stack of $D$-branes. The resulting
decoration of each open string with an assignation of Chan-Paton
factors provides a natural way to \emph{dynamically} color the ribbon
graph $\Gamma $ with labels proper of the chosen gauge group, hence
constructing out of $\Gamma $ a genuine 't Hooft diagram. Moreover, if
we consider toroidal compactifications for the target space of the
bosonic scalar fields, the $D$--branes provide an explicit expression
for the formal rules describing the $\Gamma $--interacting BCFTs on
different cylinders.  In particular, when the conformal field theory
becomes rational, we can completely characterize the dynamic of a
relevant class of fields.  These fields play a key role in the
description of the interactions between the different conformal
theories on the cylinders. We refer to them as Boundary Insertion
Operators and we provide a concrete description for both their
analytic and algebraic structure.

\subsection*{Outline}

This paper is conceptually divided into two parts. The first one,
which covers section \ref{sec:bcft}, is devoted to the construction of
a formal amplitude on the discrete open surface $M_\partial$. This is
achieved coupling such a geometry with a scalar conformal field theory,
and it involves the definition of Boundary Insertion Operators as
mediators along the interaction pattern defined by the ribbon graph $\Gamma $ associated with
$M_\partial$.

In the second part we provide an explicit prescription to dynamically
couple the above geometry with an open string gauge theory in target
space (see section \ref{sec:osgauge}). In this framework, in section
\ref{sec:ratbcft}, we provide an explicit characterization of the BCFT
interaction scheme by discussing the associated amplitude.

\section{Boundary Conformal field theory on $M_\partial$}
\label{sec:bcft}

As a starting point let us summarize the properties of the specific
geometric setup we shall deal with throughout the whole paper.  Let
$M$ denote a closed 2-dimensional oriented manifold of genus $g$. A
random Regge triangulation of $M$ is an homeomorphism
$|T_{l}|\rightarrow {M}$ where $T_{l}$ denote a $2$-dimensional
semi-simplicial complex with underlying polyhedron $|T_{l}|$ and where
each edge $\sigma ^{1}(h,j)$ of $T_{l}$ is realized by a rectilinear
simplex of variable length $l(h,j)$. Note that the connectivity of
$T_{l}$ is not a priori fixed as in the case of standard Regge
triangulations. Let $ N_{i}(T_{l})\in \mathbb{N}$ denote the number of
$i$-dimensional subsimplices $\sigma ^{i}(...)$ of $T_{l}$. Consider
the (first) barycentric subdivision $T_{l}^{(1)}$ of $
|T_{l}|\rightarrow {M}$. The closed stars, in such a subdivision, of
the vertices of the original triangulation $|T_{l}|\rightarrow {M}$
form a collection of $2$-cells $\{\rho ^{2}(i)\}_{i=1}^{N_{0}(T_{l})}$
characterizing the \emph{conical} Regge polytope
$|P_{T_{l}}|\rightarrow {M}$ barycentrically dual to
$|T_{l}|\rightarrow {M}$. Note that here we are considering a
geometrical presentation $ |P_{T_{l}}|\rightarrow {M}$ of $P$ where
the $2$-cells $\{\rho ^{2}(i)\}_{i=1}^{N_{0}(T_{l})}$ retain the
conical geometry induced on the barycentric subdivision by the
original metric structure of $ |T_{l}|\rightarrow {M}$. This latter is
locally Euclidean everywhere except at the vertices $\sigma ^{0}$,
where the sum of the dihedral angles, $\theta (\sigma ^{2})$, of the
incident triangles $\sigma ^{2}$'s is in excess (negative curvature)
or in defect (positive curvature) with respect to the $2\pi $ flatness
constraint. The corresponding deficit angle $\varepsilon $ is defined
by $\varepsilon =2\pi -\sum_{\sigma ^{2}}\theta (\sigma ^{2})$, where
the summation is extended to all $2$ -dimensional simplices incident
on the given $\sigma ^{0}$.  The automorphism group $Aut(P_{T_l})$ of
$|P_{T_{l}}|\rightarrow {M}$, (\emph{i.e.}, the set of bijective maps
preserving the incidence relations defining the polytopal structure),
is the automorphism group of the edge refinement $\Gamma $ (see
\cite{mulase}) of the $1$-skeleton of the conical Regge polytope
$|P_{T_{l}}|\rightarrow {M}$.  Such a $\Gamma$ is the $3$-valent graph
\begin{equation}
\Gamma =\left( \{\rho ^{0}(h,j,k)\}\bigsqcup^{N_{1}(T)}\{W(h,j)\},
\{\rho^{1}(h,j)^{+}\}\bigsqcup^{N_{1}(T)}\{\rho ^{1}(h,j)^{-}\}\right) .
\label{uno}
\end{equation}
where the vertex set $\{\rho ^{0}(h,j,k)\}^{N_{2}(T)}$ is identified with
the barycenters of the triangles $\{\sigma ^{o}(h,j,k)\}^{N_{2}(T)}\in
|T_{l}|\rightarrow M$, whereas each edge $\rho ^{1}(h,j)\in \{\rho
^{1}(h,j)\}^{N_{1}(T)}$ is generated by two half-edges $\rho ^{1}(h,j)^{+}$
and $\rho ^{1}(h,j)^{-}$ joined through the barycenters $\{W(h,j)
\}^{N_{1}(T)}$ of the edges $\{\sigma ^{1}(h,j)\}$ belonging to the
original triangulation $|T_{l}|\rightarrow M$. The (counterclockwise)
orientation in the $2$-cells $\{\rho ^{2}(k)\}$ of $|P_{T_{l}}|\rightarrow {
M }$ gives rise to a cyclic ordering on the set of half-edges $\{\rho
^{1}(h,j)^{\pm }\}^{N_{1}(T)}$ incident on the vertices $\{\rho
^{0}(h,j,k)\}^{N_{2}(T)}$. According to these remarks, the
(edge-refinement of the) $1$-skeleton of $|P_{T_{l}}|\rightarrow {M}$ is a
ribbon (or fat) graph \cite{mulase}, \emph{viz.}, a graph $\Gamma $ together
with a cyclic ordering on the set of half-edges incident to each vertex of $
\Gamma$. Conversely, any ribbon graph $\Gamma $ characterizes an oriented
surface $M(\Gamma )$ with boundary possessing $\Gamma $ as a spine, (\emph{
i.e.}, the inclusion $\Gamma \hookrightarrow M(\Gamma )$ is an homotopy
equivalence). In this way (the edge-refinement of) the $1$-skeleton of a
generalized conical Regge polytope $|P_{T_{l}}|\rightarrow {M}$ is in a
one-to-one correspondence with trivalent metric ribbon graphs.

It is possible to naturally relax the singular Euclidean structure
associated with the conical polytope $|P_{T_{l}}|\rightarrow {M}$ to a
complex structure $((M;N_{0}), \mathcal{C})$. Such a relaxing is
defined by exploiting \cite{mulase} the ribbon graph $\Gamma$ (see
(\ref{uno})). Explicitly, let $\rho ^{2}(h)$, $\rho ^{2}(j)$, and
$\rho ^{2}(k)$ respectively be the two-cells $\in
|P_{T_{l}}|\rightarrow {M}$ barycentrically dual to the vertices
$\sigma ^{0}(h)$, $\sigma ^{0}(j)$, and $\sigma ^{0}(k)$ of a triangle
$\sigma ^{2}(h,j,k)\in |T_{l}|\rightarrow M$. Let us denote by $\rho
^{1}(h,j)$ and $\rho ^{1}(j,h)$, respectively, the oriented edges\ of
$\rho ^{2}(h)$ and $\rho ^{2}(j)$ defined by
\begin{equation}
\rho ^{1}(h,j)\bigsqcup \rho ^{1}(j,h)\doteq \partial \rho
^{2}(h)\bigcap_{\Gamma }\partial \rho ^{2}(j),
\end{equation}
\emph{i.e.}, the portion of the oriented boundary of \ $\Gamma $ intercepted
by the two adjacent oriented cells $\rho ^{2}(h)$ and $\rho ^{2}(j)$\ \
(thus $\rho ^{1}(h,j)\in \rho ^{2}(h)$ and $\ \rho ^{1}(j,h)\in \rho ^{2}(j)$
carry opposite orientations). Similarly, we shall denote by $\rho ^{0}(h,j,k)
$ the $3$-valent, cyclically ordered, vertex of $\Gamma $ defined by 
\begin{equation}
\rho ^{0}(h,j,k)\doteq \partial \rho ^{2}(h)\bigcap_{\Gamma }\partial \rho
^{2}(j)\bigcap_{\Gamma }\partial \rho ^{2}(k).
\end{equation}

To the edge $\rho ^{1}(h,j)$ of $
\rho ^{2}(h)$ we associate \cite{mulase} a complex coordinate $
z(h,j)$ defined in the strip 
\begin{equation}
U_{\rho ^{1}(h,j)}\doteq \{z(h,j)\in \mathbb{C}|0<{Re}z(h,j)<L(h,j)\},
\label{nove}
\end{equation}
$L(h,j)$ being the length of the edge considered. The 
coordinate $w(h,j,k)$, corresponding to the $3$-valent vertex $\rho
^{0}(h,j,k)\in \rho ^{2}(h)$, is defined in the open set 
\begin{equation}
U_{\rho ^{0}(h,j,k)}\doteq \{w(h,j,k)\in \mathbb{C}\;|\;|w(h,j,k)|<\delta ,\;w(h,j,k)[\rho
^{0}(h,j,k)]=0\},
\end{equation}
where $\delta >0$ is a suitably small constant. Finally, the generic two-cell $\rho
^{2}(k)$ is parametrized in the unit disk 
\begin{equation}
U_{\rho ^{2}(k)}\doteq \{\zeta (k)\in \mathbb{C}\;|\;|\zeta (k)|<1,\;\zeta
(k)[\sigma ^{0}(k)]=0\},
\end{equation}
where $\sigma ^{0}(k)$ is the vertex $\in |T_{l}|\rightarrow M$
corresponding to the given two-cell.

\noindent We define the complex structure $((M;N_{0}),\mathcal{C})$ by
coherently gluing, along the pattern associated with the ribbon graph $
\Gamma $, the local coordinate neighborhoods $\{U_{\rho ^{0}(h,j,k)}\}_{(h,j,k)}^{N_{2}(T)}$, 
$\{U_{\rho ^{1}(h,j)}\}_{(h,j)}^{N_{1}(T)}$, and $\{U_{\rho
^{2}(k)}\}_{(k)}^{N_{0}(T)}$. Explicitly, let  $\{U_{\rho ^{1}(h,j)}\}$, $\{U_{\rho ^{1}(j,k)}\}$, $
\{U_{\rho ^{1}(k,h)}\}\ $ be the three generic open strips associated with
the three cyclically oriented edges $\rho ^{1}(h,j)$, $\rho ^{1}(j,k)$, $
\rho ^{1}(k,h)$ incident on the vertex $\rho ^{0}(h,j,k)$. Then the
corresponding coordinates $z(h,j)$, $z(j,k)$, and $z(k,h)
$\ are related to $w(h,j,k)$ by the transition functions 
\begin{equation}
w(h,j,k)=\left\{ 
\begin{tabular}{l}
$z(h,j)^{\frac{2}{3}},$ \\ 
$e^{\frac{2\pi }{3}\sqrt{-1}}z(j,k)^{\frac{2}{3}},$ \\ 
$e^{^{\frac{4\pi }{3}\sqrt{-1}}}z(k,h)^{\frac{2}{3}},$
\end{tabular}
\right. .  \label{glue1}
\end{equation}
Similarly, if $\{U_{\rho ^{1}(h,j_{\beta })}\}$, $\beta =1,2,...,q(k)$ are
the open strips associated with the $q(k)$ (oriented) edges $\{\rho
^{1}(h,j_{\beta })\}$ boundary of the generic polygonal cell $\rho ^{2}(h)$,
then the transition functions between the corresponding 
coordinate $\zeta (h)$ and each $\{z(h,j_{\beta })\}$ are given by \cite
{mulase} 
\begin{equation}
\zeta (h)=\exp \left( \frac{2\pi \sqrt{-1}}{L(h)}
\left( \sum_{\beta =1}^{\nu -1}L(h,j_{\beta })+z(h,j_{\nu })\right)
\right) ,\hspace{0.2in}\nu =1,...,q(h),  \label{glue2}
\end{equation}
with $\sum_{\beta =1}^{\nu -1}\cdot \doteq 0$, for $\nu =1$, and where
$L(h)$ denotes the perimeter of $\partial (\rho ^{2}(h))$. Iterating
such a construction for each vertex $\{\rho ^{0}(h,j,k)\}$ in the
conical polytope $|P_{T_{l}}|\rightarrow {M}$ we get a very explicit
characterization of $((M;N_{0}),\mathcal{C})$.

Such a construction has a natural converse which allows us to describe
the conical Regge polytope $|P_{T_{l}}|\rightarrow {M}$ as a
uniformization of $((M;N_{0}),\mathcal{C})$. In this connection, the
basic observation is that, in the complex coordinates introduced
above, the ribbon graph $\Gamma $ naturally corresponds to a
Jenkins-Strebel quadratic differential $\phi $ with a canonical local
structure which is given by \cite{mulase}
\begin{equation}
\phi \doteq \left\{ 
\begin{tabular}{l}
$\phi (h)|_{\rho ^{1}(h)}=dz(h)\otimes dz(h),$ \\ 
\\ 
$\phi (j)|_{\rho ^{0}(j)}=\frac{9}{4}w(j)dw(j)\otimes dw(j),$ \\ 
\\ 
$\phi (k)|_{\rho ^{2}(k)}=-\frac{\left[ L(k)\right]
^{2}}{4\pi ^{2}\zeta ^{2}(k)}d\zeta (k)\otimes d\zeta (k),$
\end{tabular}
\right.   \label{differ}
\end{equation}
where $L(k)$ denotes the perimeter of $\partial (\rho ^{2}(k))$, and
where $\rho ^{0}(h,j,k)$, $\rho ^{1}(h,j)$, $\rho ^{2}(k)$ run over
the set of vertices, edges, and $2$-cells of $|P_{T_l}|\rightarrow M$.
If we denote by
\begin{equation}
\Delta _{k}^{\ast }\doteq \{\zeta (k)\in \mathbb{C}|\;0<|\zeta (k)|<1\},
\label{puncdisk}
\end{equation}
the punctured disk $\Delta _{k}^{\ast}\subset U_{\rho ^{2}(k)}$, then
for each given deficit angle $\varepsilon (k)=2\pi -\theta (k)$ we can
introduce on each $\Delta _{k}^{\ast}$ the conical metric
\begin{eqnarray}
ds_{(k)}^{2} &\doteq &\frac{\left[ L(k)\right] ^{2}}{
4\pi ^{2}}\left| \zeta (k)\right| ^{-2\left( \frac{\varepsilon (k)}{2\pi }
\right) }\left| d\zeta (k)\right| ^{2}=  \label{metrica} \\
&=& \left|
\zeta (k)\right| ^{2\left( \frac{\theta (k)}{2\pi }\right) }
|\phi (k)_{\rho ^{2}(k)}|,  \nonumber
\end{eqnarray}
where
\begin{equation}
|\phi (k)_{\rho ^{2}(k)}|=\frac{\left[ L(k)\right]
^{2} }{4\pi ^{2}|\zeta (k)|^{2}}|d\zeta (k)|^{2}.  \label{flmetr}
\end{equation}
is the standard cylindrical metric associated with the quadratic
differential $\phi (k)_{\rho ^{2}(k)}$. Thus, the punctured Riemann
surface $((M;N_{0}),\mathcal{C})$ associated with the conical Regge
polytope $|P_{T_l}|\rightarrow M$ is provided by
\begin{gather}
((M;N_{0}),\mathcal{C});\{ds_{(k)}^{2}\})= \\
=\bigcup_{\{\rho ^{0}(h,j,k)\}}^{N_{2}(T)}U_{\rho ^{0}(h,j,k)}\bigcup_{\{\rho
^{1}(h,j)\}}^{N_{1}(T)}U_{\rho ^{1}(h,j)}\bigcup_{\{\rho
^{2}(k)\}}^{N_{0}(T)}(\Delta _{k}^{\ast },ds_{(k)}^{2}).  \nonumber
\end{gather}
Although the above correspondence between conical Regge polytopes and
punctured Riemann surfaces is rather natural there is yet another
uniformization representation of $|P_{T_l}|\rightarrow M$ which is of
relevance while discussing conformal field theory on a given
$|P_{T_l}|\rightarrow M$. The point is that the analysis of a CFT on a
singular surface such as $|P_{T_l}|\rightarrow M$ calls for the
imposition of suitable boundary conditions in order to take into
account the conical singularities of the underlying Riemann surface
$((M;N_{0}),\mathcal{C}, ds^2_{(k)})$. This is a rather delicate issue
since conical metrics give rise to difficult technical problems in
discussing the glueing properties of the resulting conformal fields.
In boundary conformal field theory, problems of this sort are taken
care of by tacitly assuming that a neighborhood of the possible
boundaries is endowed with a cylindrical metric. In our setting such a
prescription naturally calls into play the metric associated with the
quadratic differential $\phi $, and requires that we regularize into
finite cylindrical ends the cones $(\Delta _{k}^{\ast
},ds_{(k)}^{2})$.\ \ Such a regularization is realized by noticing
that if we introduce the annulus
\begin{equation}
\Delta _{\theta (k)}^{\ast }\doteq \left\{ \zeta (k)\in \mathbb{C}|e^{-\frac{
2\pi }{\theta (k)}}\leq |\zeta (k)|\leq 1\right\}\subset \overline{U_{\rho ^{2}(k)}},
\end{equation}
then the surface with boundary 
\begin{equation}
M_{\partial }\doteq ((M_{\partial };N_{0}),\mathcal{C})=\bigcup U_{\rho
^{0}(j)}\bigcup U_{\rho ^{1}(h)}\bigcup (\Delta _{\theta (k)}^{\ast },\phi
(k))
\end{equation}
defines the blowing up of the conical geometry of \ $((M;N_{0}),\mathcal{C}
,ds_{(k)}^{2})$ along the ribbon graph $\Gamma $.

The metrical geometry of \ $(\Delta _{\theta (k)}^{\ast },\phi (k))
$ is that of a flat cylinder with a circumference of length given by $L(k)$ and height given by $L(k)/\theta (k)$, (this latter being the slant radius of the
generalized Euclidean cone $(\Delta _{k}^{\ast },ds_{(k)}^{2})$ of base circumference $L(k)$ and vertex
conical angle $\theta (k)$).We also have \ \ 
\begin{eqnarray}
\partial M_{\partial } &=&\bigsqcup_{k=1}^{N_{0}}S_{\theta (k)}^{(+)}, \\
\partial \Gamma  &=&\bigsqcup_{k=1}^{N_{0}}S_{\theta (k)}^{(-)}  \notag
\end{eqnarray}
where the circles 
\begin{eqnarray}
S_{\theta (k)}^{(+)} &\doteq &\left\{ \zeta (k)\in \mathbb{C}||\zeta
(k)|=e^{-\frac{2\pi }{\theta (k)}}\right\} , \\
S_{\theta (k)}^{(-)} &\doteq &\left\{ \zeta (k)\in \mathbb{C}||\zeta
(k)|=1\right\}   \notag
\end{eqnarray}
respectively denote the inner and the outer boundary of the annulus $\Delta
_{\theta (k)}^{\ast }$. 
Note that by collapsing $S_{\theta (k)}^{(+)}$ to a point we get back the original cones $(\Delta _{k}^{\ast },ds_{(k)}^{2})$.
 Thus, the surface with boundary $
M_{\partial }$ naturally corresponds to the ribbon graph $\Gamma $
associated with the 1-skeleton $K_{1}(|P_{T_{l}}|\rightarrow {M})$ of the
polytope $|P_{T_{l}}|\rightarrow {M}$, decorated with the finite
cylinders $\{\Delta _{\theta (k)}^{\ast },|\phi (k)|\}$. In such a
framework the conical angles $\{\theta (k)=2\pi -\varepsilon (k)\}$ appears
as (reciprocal of) the moduli $m_{k}$ of the annuli $\{\Delta _{\theta
(k)}^{\ast }\}$, 
\begin{equation}
m(k)=\frac{1}{2\pi }\ln \frac{1}{e^{-\frac{2\pi }{\theta (k)}}}=\frac{1}{\theta (k)}
\end{equation}
(recall that the modulus of an annulus $r_{0}<|\zeta |<r_{1}$ is defined by $\frac{1}{2\pi }\ln \frac{r_{1}}{r_{0}}$). According to these remarks we can equivalently represent the conical Regge polytope $|P_{T_{l}}|\rightarrow {M}$ with the uniformization  $((M;N_{0}),\mathcal{C});\{ds_{(k)}^{2}\})$ or with its blowed up version $M_{\partial }$.

In order to exploit the above geometrical set up in the study of
open/closed string dualities, let us consider $D$ real scalar maps
$X^\alpha:\,M_\partial \rightarrow \mathcal{T},\,i = 0, \ldots, D-1$,
injecting $M_\partial$ into an unspecified target space $\mathcal{T}$
and let us first focus on a fixed, but otherwise generic, \cyl{p}.
Although quantization of (non-critical) Polyakov string on a annular
domain is an overkilled topic, it is worthwhile discussing it in some
detail, both to fix notation and to deal with some of the subtleties
arising from to the combinatorial origin of $M_\partial$.

The world-sheet action is: 
\begin{multline}
  \label{eq:action}
  S
\,=\,
\frac{1}{4\pi}
\int d \zeta(p) d \bgz(p)\, 
G_{\alpha \beta}(p) \partial X^\alpha(p)
\bar{\partial} \overline{X}^\beta(p) \,+\,\\
B_{\alpha \beta}(p) \partial X^\alpha(p)
\bar{\partial} \overline{X}^\beta(p)  \,-\,
\frac{1}{2} \Phi(p)\,R^{(2)}.
\end{multline}

The geometry of target space is specified by a suitable assignation of
the background matrix
\begin{equation}
  \label{eq:bm}
  E(p) \,=\, G(p) \,+\, B(p),
\end{equation}
which encodes informations about the background metric $G_{\alpha
  \beta}(p)$ and the Kalb-Ramond field $B_{\alpha \beta}(p)$
components.  $\Phi(p)$ is a properly chosen dilaton field. In
particular, we will deal with flat toroidal backgrounds, \ie{} we will
consider a string moving in a background in which $D$ dimensions are
compactified whereas the metric, the Kalb-Ramond field and the dilaton
are independent from the spacetime coordinates $X^\alpha, \alpha = 1,
\ldots, D$.

Since in the description of the metric geometry of the triangulation
$|T_{l}|\rightarrow M$ as the dual open Riemann surface $M_\partial$
we are, roughly speaking, unwrapping conical 2-cells into finite
cylindrical ends \cite{Carfora1}, we can adopt for the matter sector the
most general condition:
\begin{equation}
  \label{winding}
  X^\alpha(p)(e^{2 \pi i}\zeta,\,e^{-2 \pi i}\overline{\zeta}) \,=\, 
  X^\alpha(p)(\zeta,\,\overline{\zeta}) 
  \,+\, 2 \nu^\alpha(p) \pi \frac{R^\alpha(p)}{l(p)}, \qquad \quad 
  \nu^\alpha(p)\,\in\,\mathbb{Z}
\end{equation} 
according to which each field $X^\alpha(p)$ winds $\nu^\alpha(p)$ times
around the corresponding toroidal cycles of length $\frac{R^\alpha(p)}
{l(p)}$ in the compact target space $\mathcal{T}$. Here $l(p)$ is a 
length parameter built out of the geometric assigned data of the 
original triangulation.

In this way, if we further put to zero the dilaton and $B$-field
components, we are actually encoding all data about the background
geometry in the value of the compactification radius, letting the
metric to be diagonal and decoupling the model in each direction.
Hence, we can consider just the quantization of a single
scalar field. The world-sheet action on \cyl{k} becomes:
\begin{equation}
  \label{eq:action1}
  S\,=\,
  \frac{1}{8\pi} \int\limits_{\Delta^*_{\epsilon(p)}} d\zeta(p) d\bgz(p)\,
  \partial X(p)\, \overline{\partial} \overline{X}(p).  
\end{equation}

The extension to a $D$-dimensional background will be straightforward
from a target-space point of view.

Since the theories on the various cylindrical ends are effectively decoupled,
from now on we shall suppress the polytope index $(k)$ and we will
restore it once we will describe the interaction of the distinct
models along the ribbon graph $\Gamma$.

The fundamental prerequisite to quantize a Conformal Field Theory
(CFT) on a surface with boundary is to have the full control of the
same quantum theory on the entire complex plane, the latter being
usually referred to as the \emph{bulk theory}. This is defined via a
suitable assignation of an Hilbert space of states $\mathcal{H}^{(C)}$,
endowed with the action of an Hamiltonian operator $H^{(C)}$ and of a
vertex operation, \ie{} a formal map
$\Phi^{(C)}(\circ;\,\zeta,\,\bgz): \; \mathcal{H}^{(C)}
\,\rightarrow\, \text{End}\left[V [\zeta,\bgz]\right]$ associating to
each vector $\vert\phi\rangle \in \mathcal{H}^{(C)}$ a conformal field
$\phi(\zeta,\,\bgz)$ of conformal dimension $h,\bar{h}$.  The bulk
theory is completely worked out once we know the coefficients of the
Operator Product Expansion (OPE) for all fields in the theory.
Actually, we can face this task for most CFTs since, among conformal
fields, a preferential role is played by chiral ones, whose Laurent
modes generate two isomorphic copies of the chiral algebra which
defines the symmetries of the theory.

In our case such a role is played by chiral currents $ J(z) \,=\,
i\,\partial X(z) \,=\, \sum_n \fa_n \, z^{-n-1}$ and
$\overline{J}(\bgz) \,=\, i\,\overline{\partial} \overline{X}(\bgz)
\,=\, \sum_n \ofa_n \, \bgz^{-n-1} $ which generate two independent
copies of the Heisenberg algebra:
\begin{gather}
  \label{heis}
  \left[ \fa_n \,,\, \fa_m \right] \,=\, 
  n \delta_{n+m,0}
  \qquad
  \left[ \ofa_n \,,\, \ofa_m \right] \,=\, 
  n \delta_{n+m,0}  
  \notag \\
  \left[ \fa_n \,,\, \ofa_m \right] \,=\, 0.
\end{gather}

The Virasoro fields $T$ and $\overline{T}$ play a special role among
the chiral fields of a CFT. Their modes $L_n$ and $\overline{L}_n$
close two copies of the Virasoro algebra:
\begin{gather*}
  \left[L_m\,,\,L_n\right] 
  \,=\, (m - n)\,
  L_{m + n} \,+\, \frac{1}{12} n (n^2 - 1) \delta_{m + n , 0},\\
  \left[\overline{L}_m\,,\,\overline{L}_n\right] 
  \,=\, (m - n)\,
  \overline{L}_{m + n} \,+\, \frac{1}{12} n (n^2 - 1) \delta_{m + n ,
  0}.
\end{gather*}

Since the Virasoro algebra belongs to the universal covering of the
Virasoro one, we can represent its generators by means of the Sugawara
construction:
\begin{equation}
  \label{virasoro}
  L_0
  \,\doteq\, 
  \sum_{n>0}
  \mathfrak{a}_{-n} \cdot
  \mathfrak{a}_n 
  \,+\,
  \frac{1}{2}(\mathfrak{a}_0)^2, 
  \qquad
  L_n
  \,\doteq\, 
  \frac{1}{2}\sum_{m \in \mathbb{Z}}
  \ordprod{\mathfrak{a}_{n - m}\cdot \mathfrak{a}_m}, 
\end{equation}
hence allowing for an immediate definition of the bulk
Hamiltonian operator $H^{(C)}$:
\begin{equation*}
  H^{(C)}  \,=\,
  \frac{2\pi}{L} 
  \left( 
    L_0 \,+\, \bar{L}_0 \,-\, \frac{D}{12}
  \right).
\end{equation*}
Moreover, their action determines a diagonal decomposition of the
Hilbert space into subspaces carrying irreducible representations of the 
two commuting chiral algebras:
\begin{equation}
  \label{eq:chirdec}
  \mathcal{H}^{(C)} \,\doteq\,
  \bigoplus_{\lambda\,\overline{\lambda}}
  \mathcal{H}_\lambda
  \,\otimes\, 
  \overline{\mathcal{H}}_{\overline{\lambda}},
\end{equation}
where $\lambda_{(\mu,\nu)} \,=\, \mu \frac{l}{R} \,+\, \frac{1}{2} \nu
\frac{R}{l}$ and $\overline{\lambda}_{(\mu,\nu)} \,=\, \mu \frac{l}{R}
\,-\, \frac{1}{2} \nu \frac{R}{l}$ are respectively the 
$U(1)_L$ and $U(1)_R$ charges (real numbers).

\subsection{Amplitude on \cyl{p}}

\begin{figure}[!t]
  \centering
  \includegraphics[width=.7\textwidth]{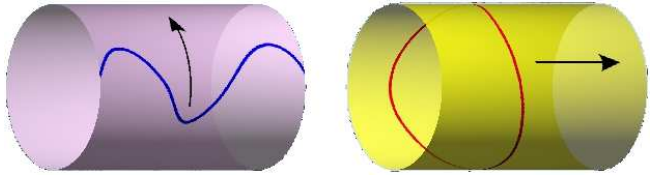}
  \caption{Dual cylinders.}
  \label{fig:dual_cyl}
\end{figure}

The bulk CFT's properties we briefly summarized in last section are
the main ingredients to discuss in detail the extension of the same
CFT on a given cylindrical end over $M_\partial$. As a matter of fact,
remembering that, from a microscopic point of view, to define a CFT on
a surface with boundary means to work out  which values we can consistently
assign to fields on the boundaries of the new domain (\ie{} which
boundary conditions we can choose), the key datum we must keep track
of to fulfill this goal are the OPE coefficients of the bulk
theory. The latter identify an algebra of fields the boundary
assignations must be compatible with. Thus, the recipe we will follow
aims, firstly, to look for all possible boundary assignations. We will
show that these can be encoded into a set of coherent boundary states
which arise as a generalization of those introduced in
\cite{Langlands}. Secondly, we will consider the list of constraints
which the bulk algebra of fields induces on the boundary components to
select, among the above set of boundary assignations, those compatible
with the algebra itself.

Thus, let us consider any but fixed cylindrical end \cyl{k}. In a
string theory perspective, it can be viewed both as an open string
one-loop diagram or as the tree level diagram of a closed string
propagating for a finite length path.  In the first (direct channel)
picture, time flows around the cylinder and the associated
quantization scheme defines functions of the modular parameter
$\tau(p) = i \theta(p) = i (2 \pi - \varepsilon(p))$.  On the
opposite, in the second (transverse channel) framework, time flows
along the cylinder, and the associated quantization scheme is related
to the former by means of the modular transformation $\tau(p)
\rightarrow -\frac{1}{\tau(p)}$. In the forthcoming analysis, we will
switch back and forth between these two points of view.

The key object we wish to calculate is the amplitude associated to
this diagram; this is a deep-investigated problem whenever the
boundary assignations a priori satisfy the canonical prescription of
Neumann or Dirichlet conditions \cite{Saleur:1998hq}.  However, within
the discretized model we are dealing with, cylindrical ends arise as a
byproduct of an unwrapping process of a conical structure
\cite{Carfora1, Carfora3}. Hence we do not have a priori a full
control on the behavior of matter fields on the vertexes of the parent
triangulation when these spread over the full outer cylinder
boundary\footnote{Within this manuscript, we keep the convention of
  \cite{Carfora1, Carfora3} to refer to $S^{(+)}_{\varepsilon(p)}$ as
  the boundary of the cylinder glued to the ribbon graph (inner
  boundary in the annuli picture) whereas $S^{(-)}_{\varepsilon(p)}$
  is the ``free boundary'' (outer boundary in the annuli picture).}
$\boum$. Thus, in our description, we will follow the procedure
outlined by Charpentier and Gawedzky in \cite{Charpentier:1990gn},
which allows to write the amplitude on an arbitrary Riemann surface
$\Sigma$ with a fixed number of boundary loops $S_I$, parametrized by
analytical real maps $p_I : S^1 \rightarrow S_I$, and by an arbitrary
specification of matter fields on them. 

Within this framework we get:
\begin{equation}
  \label{eq:form_ampl}
\mathcal{A}_\Sigma \,=\,
  \int_{\{ X \circ p_I \,=\, X_I\}} \mathcal{D} \left[X\right]\,
  e^{- S[X]}, 
\end{equation}
where $S[X]$ is the Euclidean action of the bulk CFT,
$\mathcal{D}\left[ X\right]$ is the formal measure on the target space
and the kinematical configurations of the field $X$ are those such
that it assumes the general but fixed value $X_I$ over the given
boundary loop $S_I$.

This rather abstract and formal expression acquires a precise meaning
when we deal with real scalar fields defined as injection maps from
$\Sigma$ to a flat toroidal background.  In this case, it is always
possible to decompose $X = X_{cl} + \tilde{X}$ where the real map
$X_{cl}$ is an harmonic function w.r.t. $\Delta_\Sigma$ (the Laplacian
operator defined over $\Sigma$) fulfilling the boundary assignation
(\ie{} $ X_{cl} \circ p_I \,=\, X_I$).
$\tilde{X}\,:\;\Sigma\rightarrow\mathbb{R}$ is the collection of the
off-shell modes of $X$ satisfying $\tilde{X} \circ p_I = 0$.  This
constraint implies the diagonal decomposition of the bulk action,
$S[X] \,=\, S[X_{cl}] + S[\tilde{X}]$. If we specify $\Sigma \doteq
\text{\cyl{k}}$, we get:
\begin{equation}
  \label{eq:form_ampl2}
  \mathcal{A}_{\text{\cyl{p}}} \,=\, 
  \frac{1}{4\pi} \frac{1}{\eta(\tilde{q})}
\sum_{X_{cl}} e^{- S[X_{cl}]}, 
\end{equation}
where $\eta(\tilde{q})$ is the Dedekind-$\eta$ function with
$\tilde{q} = e^{- \frac{2 \pi i}{\tau(p)}}$ and where the sum runs 
over the set of classical solutions.

According to the prescription introduced in \cite{Charpentier:1990gn}
and specialized to the compactified boson in \cite{Langlands}, it is
possible to parametrize the classical field (zero mode) in term of its
restrictions to the boundaries $\boup$ and $\boum$.  As a byproduct,
the space of classical solutions is fully parametrized by the two set
of complex numbers $\left\{a_n \right\}$ and $\left\{ b_n \right\}$,
obeying the reality conditions $a_{-n} \,=\, \overline{a}_n$ and
$b_{-n}\,=\,\overline{b}_n$, a real number
$t\,\in\,\left(0,\,2\pi\frac{R}{l}\right]$ and a pair of integers
$(\mu,\nu) \,\in\,\mathbb{Z}^2$. The latter are in one-to-one
relation with the two integers parametrizing the irreducible
representations of the chiral algebra (see eq.  \eqref{eq:chirdec} and
comments below). The existence of such 1:1 correspondence can be fully
exploited to make explicit the formal map between admissible boundary
conditions and coherent states built out of linear combinations of
elements in the bulk Fock space. In details, splitting $t = t_- -
t_+$, we map each boundary assignation labelled by
$\left\{(\mu,\nu),\,\{a_n\},\,t_-\right\}$ (resp.
$\left\{(\mu,\nu),\,\{b_n\},\,t_+\right\}$) into $\left|
  \mathfrak{r}_{(\mu ,\nu )}^{\alpha }(S_{\varepsilon
    (k)}^{(-)})\right\rangle$ (resp. $\left| \mathfrak{r}_{(\mu ,\nu
    )}^{\alpha }(S_{\varepsilon (k)}^{(+)})\right\rangle$) $\in
\mathcal{H}_{\mu,\nu} \otimes \overline{\mathcal{H}}_{\mu,\nu}\subset
\mathcal{H}^{(C)}$.

Therefore, the amplitude on the fixed cylindrical end can be written as
\begin{equation}
  \label{eq:ampl_bs}
  \mathcal{A}(\left\{ a_n \right\}\left\{
    b_n\right\}, t) \,=
  \sum_{(\mu,\nu)}
  \left\langle\mathfrak{r}_{(\mu ,\nu )}(S_{\varepsilon
      (p)}^{(+)})\right\vert 
  \tilde{q}^{L_0 \,+\, \bar{L}_0 \,-\, \frac{c}{12}} 
  \left| \mathfrak{r}_{(\mu ,\nu )}(S_{\varepsilon
      (p)}^{(-)})\right\rangle.
\end{equation}

These boundary states are the following generalization of those
introduced in \cite{Langlands}:
\begin{multline}
\label{eq:lbs-}
  \left| \mathfrak{r}_{(\mu ,\nu )}(S_{\varepsilon
      (p)}^{(-)})\right\rangle =e^{i t_{-}(\lambda _{(\mu ,\nu
      )}+\overline{\lambda }_{(\mu ,\nu )})}
  \times \\ \times
  \prod_{n=1}^{\infty
  }\sum_{m_{1},m_{2}}
  A_{m_{1},m_{2}}^{n}(a_{n},a_{-n})%
  \frac{\left( \mathfrak{a}_{-n}\right)^{m_{1}}
    \otimes
    \left( 
      \overline{\mathfrak{a}}_{-n}
    \right)%
    ^{m_{2}}}{\sqrt{n^{m_{1}+m_{2}}m_{1}!m_{2}!}}
  \left|
    (\mu,\nu)
  \right\rangle, 
\end{multline}
and
\begin{multline}
\label{eq:lbs+}
  \left| \mathfrak{r}_{(\mu ,\nu )}(S_{\varepsilon
      (p)}^{(+)})\right\rangle =e^{i t_{+}(\lambda _{(\mu ,\nu
      )}+\overline{\lambda }_{(\mu ,\nu )})}
  \times \\ \times
  \prod_{n=1}^{\infty
  }\sum_{m_{1},m_{2}}
  B_{m_{1},m_{2}}^{n}(b_{n},b_{-n})%
  \frac{\left( \mathfrak{a}_{-n}\right)^{m_{1}}
    \otimes
    \left( 
      \overline{\mathfrak{a}}_{-n}
    \right)^{m_{2}}}{\sqrt{n^{m_{1}+m_{2}}m_{1}!m_{2}!}}
  \left|
    (\mu,\nu)
  \right\rangle,
\end{multline}
with
\begin{multline}
  A_{m_{1},m_{2}}^{n}(a_{n},a_{-n})
  \,=\, e^{i \pi s(m_1 + m_2)} \times\\
  \begin{cases}
    (2i\sqrt{n}a_{n})^{m_{1}-m_{2}}\sqrt{\frac{m_{2}!}{m_{1}!}}%
    e^{-2n|a_{n}|^{2}}L_{m_{2}}^{(m_{1}-m_{2})}(4n|a_{n}|^{2}), 
    & m_{1}\geq m_{2} \\
    (2i\sqrt{n}a_{n})^{m_{2}-m_{1}}\sqrt{\frac{m_{1}!}{m_{2}!}}%
    e^{-2n|a_{n}|^{2}}L_{m_{1}}^{(m_{2}-m_{1})}(4n|a_{n}|^{2}), 
    & m_{2}\geq m_{1}
  \end{cases}
\end{multline}
being $s\in\mathbb{R}$ and $L_{m_{2}}^{(m_{1}-m_{2})}(\circ )$ the 
$m_{2}$-th Laguerre
polynomial of the $(m_{1}-m_{2})$-th kind.  Replacing $a_{n}$
with $b_{n}$ and acting by conjugation (induced by the orientation of
the boundary), we end up with
$B_{m_{1},m_{2}}^{n}(b_{n},b_{-n}) =
\overline{A_{m_{1},m_{2}}^{n}(b_{-n},b_{n})}$. We leave a
reader interested in the precise derivation to \cite{gili-phd}.

Although exhaustive from a mathematical point of view, as anticipated
at the beginning of the section the answer we reached with
\eqref{eq:lbs-} and \eqref{eq:lbs+} is not yet conclusive. As a matter
of fact, from a physical point of view, the presence of a boundary
allows us to rephrase the whole process macroscopically considering
the presence of two branes which, in the transverse channel, emit and
absorb a closed string (whose initial and final states are described
by the above boundary states) via non-perturbative processes, while,
in the direct channel they are the objects where the endpoints of the
open string running one loop lay on.  In this connection, BCFT is the
natural mean to describe microscopically the brane-string bound state,
without any reference to spacetime geometry. As a consequence, we need
to avoid any information flow through the boundary itself (the
cylinder or the annulus boundary in our setting) and, to this avail,
chiral and Virasoro fields must satisfy appropriate glueing conditions
along it. In particular, the holomorphic and antiholomorphic
components of the latter must coincide on the annulus boundary:
\begin{subequations}
\label{eq:uguali?}
  \begin{equation}
    \label{eq:uguali}
    \zeta^2\, T(\zeta)_{|\zeta|=\q}\,=\,
\bgz^2\bar{T}(\bgz)|_{|\zeta|=\q} \quad\text{and}\quad \zeta^2\, T(\zeta)_{|\zeta|=
  1} \,=\,\bgz^2\bar{T}(\bgz)|_{|\zeta|= 1}.
  \end{equation}
The analogue condition on the chiral currents is weaker; they must be
related by an automorphism $\Omega$ of the chiral algebra:
\begin{equation}
  \label{eq:ug-auto}
\zeta\,
J(\zeta)_{|\zeta| =\q}\,=\, \bgz\Omega\bar{J}(\bgz)|_{|\zeta|=\q}
\quad\text{and}\quad 
 \zeta\, J(\zeta)_{|\zeta| =1}\,=\,
\bgz\Omega\bar{J}(\bgz)|_{|\zeta|=1}.  
\end{equation}
\end{subequations}

Being the $\mathfrak{u}(1)$
algebra Abelian, its automorphism group is $\mathbb{Z}_2$, thus
$\Omega = \pm 1$. Exploiting radial quantization the above glueing
conditions translate into projection maps acting on boundary states:
  \begin{equation}
    \label{cvir}
  \left( L_n \,-\, \overline{L}_{-n} \right) \,
  \Vert B \Rangle \,=\, 0,    
\end{equation}
and 
\begin{subequations}
  \label{cond}
\begin{gather}
  \label{cpiu}
  \left( \mathfrak{a}_n \,+\, \overline{\mathfrak{a}}_{-n} \right) \,
  \Vert B \Rangle \,=\, 0,  \qquad \text{if}\: \Omega\,=\,-1 \\
  \label{cmeno}
  \left( \mathfrak{a}_n \,-\, \overline{\mathfrak{a}}_{-n} \right) \,
    \Vert B \Rangle \,=\, 0. \qquad \text{if}\: \Omega\,=\,+1.
  \end{gather}
\end{subequations}

The Sugawara construction ensures that conditions \eqref{cpiu} and
\eqref{cmeno} are
sufficient to enforce conformal invariance encoded in \eqref{cvir}.
Their application projects \eqref{eq:lbs-} and \eqref{eq:lbs+} into the 
ordinary Neumann and Dirichlet boundary states defined for the
compactified bosonic field $X(\zeta,\bgz)$:
\begin{subequations}
\begin{gather}\label{eq:Cla3}
  \left| \mathfrak{r}_{(\mu ,\nu )}(S_{\varepsilon
      (p)}^{(-)})\right\rangle ^{(D)} 
  =
  \frac{1}{\sqrt{2 \frac{L(p)}{R(p)}}} 
  \sum_{\mu \in \mathbb{Z}}
  e^{i t_{+}\mu
    \frac{L(p)}{R(p)}}e^{\sum_{n=1}^{\infty }\frac{1}{n}%
    \mathfrak{a}_{-n}\overline{\mathfrak{a}}_{-n}} 
  \left| (\mu,0)\right\rangle,
  \\
  \left| \mathfrak{r}_{(\mu ,\nu )}(S_{\varepsilon
      (p)}^{(-)})\right\rangle ^{(N)}
  =
  \sqrt{\frac{L(p)}{R(p)}}
  \sum_{\nu \in \mathbb{Z}}
  e^{i \tilde{t}_{+}\frac{\nu}{2}
    \frac{R(p)}{L(p)}}
  e^{-\,\sum_{n=1}^{\infty }\frac{1}{n}
    \mathfrak{a}_{-n}\overline{\mathfrak{a}}_{-n}} 
  \left| (0,\nu)\right\rangle.\label{eq:Cla4}
\end{gather}
\end{subequations}
An equivalent relation clearly holds for $\left| \mathfrak{r}_{(\mu ,
\nu )}(S_{\varepsilon (p)}^{(+)})\right\rangle$.
The careful demonstration of the above statements, which exploits
recursion relations of the Laguerre polynomials, is reported in
\cite{gili-phd}, section 2.3 and appendix A.

\subsection{Interactions on $\Gamma$: Boundary Insertion Operators}
\label{sec:bio}

With the previous analysis we determined the set of boundary states
representing the admissible field assignations over each \cyl{k}
boundary component. This not only completes the first of 
the two step programme outlined at the beginning of the section, but
it is instrumental for the next one, the discussion on
the interaction along the ribbon graph $\Gamma$ among the $N_0$
distinct copies of the theory, each one living on a different
cylindrical end.

The existence of pairwise adjacent boundary conditions led us to
propose in \cite{Carfora2} that this interaction could be mediated by
boundary conditions changing operators, whose presence is predicted in
the abstract formulation of boundary conformal field theory
\cite{Cardy1, Recknagel:1998ih}.  As a matter of fact, in 
the standard scenario of a BCFT defined on the the Upper Half Plane,
the prescribed boundary condition can change along the real axis.
In a radial quantization scheme, such a situation is explained with
the presence of a vacuum which is no longer invariant under the action
of the Virasoro operator $L_{-1}^{(H)}$. In \cite{Cardy1} it
was proposed that such states are obtained by the local action of a
specific operator acting on the true vacuum and supported only on the
boundary, \ie{} it induces a transition between boundary conditions.
According to the vertex operation, each of these operators can be
associated to a specific vector in the Fock space dependent upon
boundary data and such that it cannot be correlated with bulk fields by 
means of a bulk to boundary OPE.

\begin{figure}[!t]
  \begin{center}
    \includegraphics[width=.35\textwidth]{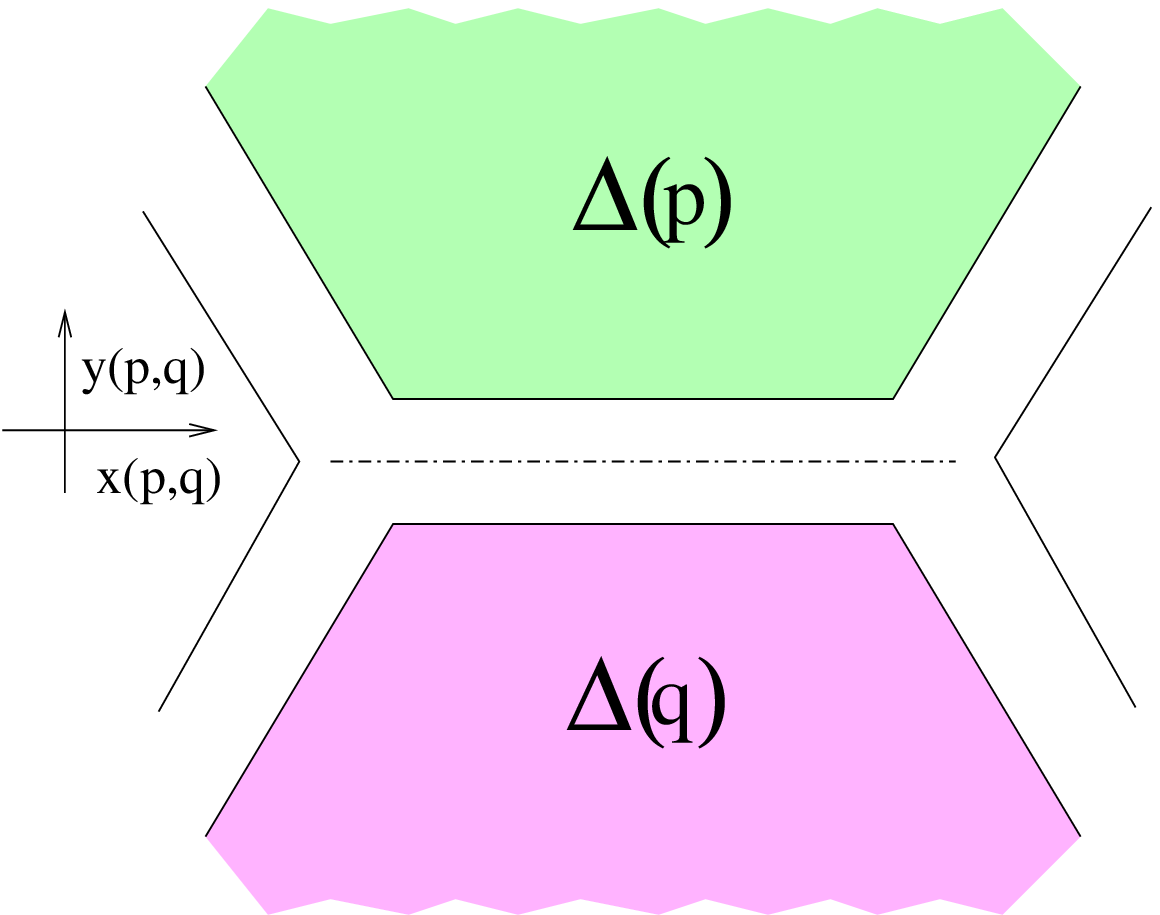}
    \quad
    \raisebox{1.5cm}{$\xrightarrow[y(p,q) \rightarrow 0]
      {\text{ADJACENCY LIMIT}}$}
    \quad
    \includegraphics[width=.3\textwidth]{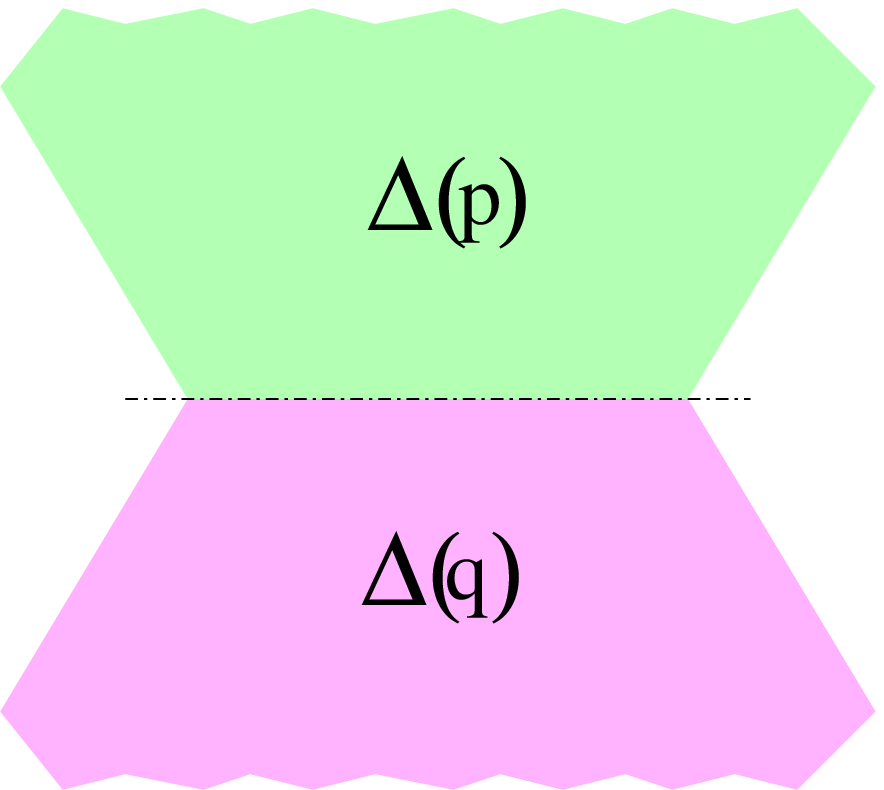}
    \caption{Shared boundaries in the adjacency limit.} 
    \label{fig:BIO}
  \end{center}
\end{figure}

However, the described local action of a boundary condition changing
operator does not fit in our discretized model. As a matter of fact,
in the framework dual to a Random Regge Triangulation, the $N_0$ cylinders
are pairwise glued together along one of their two boundaries (commonly the inner
one in the annuli picture) through one ribbon graph edge. 
Hence, in this case, we should more
properly speak of a ``separation edge''\footnote{In order to keep
  terminology and notations ``under control'', we shall often refer to
  the edge $\rho^1(p,q)$ in common between \cyl{p} and \cyl{q} as a
  boundary. We feel that the overall context allows the reader to
  point out which is the specific scenario we are dealing with.}
between two adjacent cylindrical ends. Furthermore, we do not
have a jump between two boundary conditions taking place at a precise
point. On the opposite, two different boundary conditions coexist in
the adjacency limit along the whole edge \cite{Carfora2}, as depicted
in fig. \ref{fig:BIO}.

Switching back to field theoretical contents, in this connection it is
no longer correct to claim the presence of a vacuum state invariant
under translations along the boundary. As a matter of facts, being the
shared boundary obtained out of two separate loops, each of them being
part of a domain where a BCFT is constructed, all the associated Fock
space elements are invariant under translation only along the relevant
boundary loop.  Thus, in order for the geometric glueing process to be
consistent with the functional data of the theory on each cylinder, we
must require that the $N_0$ a priori independent Fock spaces blend
pairwise without breaking the conformal and the chiral symmetry of the
model.  As we will show in the forthcoming discussion, this leads to
the introduction of an additional class of operators which live on
the boundary shared between two adjacent polytopes, carry an
irreducible action of the chiral algebra and dynamically mediate
between two adjacent boundary conditions.

To provide the details, let us consider two adjacent cylindrical ends
\cyl{p} and \cyl{q}; these ends are glued to the oriented boundaries
$\partial\Gamma_p$ and $\partial\Gamma_q$ of the ribbon graph.  Let us
consider the oriented strip associated with the edge $\rho^1(p,q)$ of
the ribbon graph and its uniformized neighborhood
$\left(U_{\rho^1(p,q)},z(p,q)\right)$, where the uniformizing
coordinate $z(p,q)$ was defined as in \eqref{glue2}. In this
geometric background, let us focus on an unspecified BCFT on \cyl{p}
and let us fix some notations:
\begin{itemize}
\item $W(\zeta)$ and $\overline{W}(\bar{\zeta})$ are respectively the
  set of holomorphic and antiholomorphic chiral fields of the parent
  bulk theory defining two commuting copies of the chiral algebra
  describing the symmetries of the model;
\item $\mathcal{Y} = \{\lambda(p)\}$ is the collection of indexes
  labelling the irreducible representations of the chiral algebra
  associated to the BCFT on \cyl{p};
\item $\mathcal{A} = \{A(p)\}$ is the set of possible boundary
  conditions we can assign on each boundary components, hence located
  at $|\zeta(p)|=1$ and at $|\zeta(p)|=\q$ in the annuli picture. Each
  $A(p)$ includes either the glueing automorphism, denoted as
  $\Omega_{A(p)}$, either a specification for all other necessary
  parameters (\ie{}, when dealing with the compactified boson, the
  brane position or the Wilson line).
\end{itemize}

Beside avoiding information flow through the boundary (see comments
before formulae \eqref{eq:uguali?}), the existence of the glueing
automorphism $\Omega_{A(p)}$ cited above gives rise to the action of a
single copy of the chiral algebra on the state space
$\mathcal{H}^{(O)}$ of the boundary theory. As a matter of fact, being
defined only on a part of the full complex plane {\it i.e.} the
annulus, $W(\zeta)$ and $\bar{W}(\bgz)$ are not sufficient to generate
two copies of the chiral algebra. However, since the glueing condition
$ \zeta(p)^{h_W} W(\zeta(p))_{|\zeta(p)|=1} = \Omega_{A(p)}
\zeta(p)^{\bar{h}_{\overline{W}}}
\overline{W}(\bgz(p))|_{|\zeta(p)|=1}$ (being $h_W$ the conformal
weight of $W(\zeta(p)$) states that holomorphic and antiholomorphic
chiral fields are related on the boundary, we may
introduce\footnote{The reader should keep track of the following
  change of perspective: $\zeta(p)$ and $\bar{\zeta}(p)$ are no more
  independent coordinates but, in formulas such as \eqref{eq:1al} they
  are related by complex conjugation.}:
\begin{equation}
  \label{eq:1al}
  \mathbf{W}_{\Omega_{A(p)}} = 
  \begin{cases}
    W(\zeta(p)) & 
    |\zeta(p)| \leq 1 \\
    \Omega_{A(p)} \overline{W}(\bgz(p)) & 
    |\zeta(p)| > 1 \\
  \end{cases},
\end{equation}

which is a single analytic function on $\mathbb{C}$.  Its Laurent
expansion coherently defines a single copy $\mathcal{W}$ of the chiral
algebra associated to the boundary conformal field theory on \cyl{p}
\cite{Cardy1,Recknagel:1998ih}. Hence it induces a decomposition of
the \emph{open} CFT Fock space $\mathcal{H}^{(O)}$ into a sum of
carriers of its irreducible representations \cite{Gaberdiel:2002iw}:
$\mathcal{H}^{(O)} \, = \, \bigoplus_\lambda \mathcal{H}_\lambda$,
being $\mathcal{H}_\lambda$ the subspace appearing in
\eqref{eq:chirdec}.

The above construction and discussion holds for the 
BCFT defined each cylinder. Suppose now to held fixed \cyl{p} and let
us consider its adjacent cylinder \cyl{q}. Referring to $B(q)$ as the 
boundary condition on its inner boundary out of the automorphism 
$\Omega_{B(q)}$, the glueing condition reads 
$$\zeta(q)^{h_W} W(\zeta(q))_{|\zeta(q)|=1} = \Omega_{B(q)}
\bgz(q)^{\bar{h}_{\overline{W}}} \overline{W}(\bgz(q))|_{|\zeta(q)|=1},$$
whereas the single chiral field is
\begin{equation*}
  \mathbf{W}_{\Omega_{B(q)}} = 
          \begin{cases}
            W(\zeta(q)) & 
            |\zeta(q)| \leq 1 \\
            \Omega_{B(q)} \overline{W}(\bgz(q)) & 
            |\zeta(q)| > 1 \\
          \end{cases},
\end{equation*}
which is analytic on the full complex plane and whose Laurent modes
define a single copy of the chiral algebra. 

Within this framework we can implement a non symmetry-breaking glueing
of two adjacent cylindrical ends associating to such a pair a unique
copy of the chiral algebras and, by means of the Sugawara
construction, of the Virasoro ones.  To this avail, as a starting
point we exploit \eqref{glue2} to express the holomorphic and the
antiholomorphic components of the chiral fields defined on each
cylindrical end in term of the strip coordinate, namely $
\mathbf{W}_{\Omega_{A(p)}} (z(p,q)) \,=\, \mathbf{W}_{\Omega_{A(p)}}
(\zeta(p)) \left( \frac{d\,z(p,q)}{d\,\zeta(p)}
\right)^{-h_{\mathbf{W}}}$ and $ \mathbf{W}_{\Omega_{B(q)}} (z(q,p))
\,=\, \mathbf{W}_{\Omega_{B(q)}} (\zeta(q)) \left(
  \frac{d\,z(q,p)}{d\,\zeta(q)} \right)^{-h_{\mathbf{W}}}$.

Taking into account $z(q,p) = - z(p,q)$, we perform the glueing
requiring a condition similar to \eqref{eq:uguali?}
to hold. In this process, the subtle point resides in the map
$\Omega$ in \eqref{eq:1al}. As a matter of fact, we must take into
account that the whole process must relate the two glueing
automorphisms $\Omega_{A(p)}$ and $\Omega_{B(q)}$ associated to the
BCFTs defined respectively on \cyl{p} and \cyl{q}. Thus it seems
natural to introduce a further automorphism ${\Omega'}^{A(p) B(q)}$
which, in the adjacency limit $y(p,q)=\Im\left[z(p,q)\right]
\rightarrow 0$, acts along the boundary deforming continuously the
(holomorphic and antiholomorphic components of the) bulk chiral fields
in \cyl{p} in the corresponding on \cyl{q}.  To rephrase:
\begin{equation}
\label{eq:ifr}
\mathbf{W}_{\Omega_{A(p)}} (z(p,q))|_{y(p,q) \rightarrow 0} =
{\Omega'}^{A(p) B(q)}\mathbf{W}_{\Omega_{B(q)}}(z(p,q)) |_{y(p,q) 
\rightarrow 0}.  
\end{equation}

In this way, we are indeed implementing a two way dynamical flow of
informations between \cyl{p} and \cyl{q}.  As a matter of fact,
\eqref{eq:ifr} provides a concrete mean to associate to each pairwise
adjacent set of conformal theories a unique chiral current out of
\eqref{eq:ifr}:
\begin{equation}
\label{eq:pq_cur}
  \mathbb{W}_{\Omega_{A(p) B(q)}}\left(z(q,p)\right) =
  \begin{cases}
    \mathbf{W}_{\Omega_{A(p)}}\left(z(q,p)\right) & \text{in\;\cyl{p}
    $\cup\rho_1(p,q)$} \\
    {\Omega'}^{A(p) B(q)}\mathbf{W}_{\Omega_{B(q)}}\left(z(q,p)\right) &
    \text{in\;\cyl{q}$\cup\rho_1(q,p)$}
  \end{cases}.     
\end{equation}
We emphasize that, although the second component of $\mathbb{W}_{\Omega
_{A(p) B(q)}}\left(z(q,p)\right)$ should be naturally expressed in term of the
coordinate $z(p,q)$, we implicitly exploit the relation $z(q,p)=-z(p,q)$
in order to avoid an unnecessary redundancy.
 
\begin{figure}[!t]
  \centering
  \includegraphics[width=.75\textwidth]{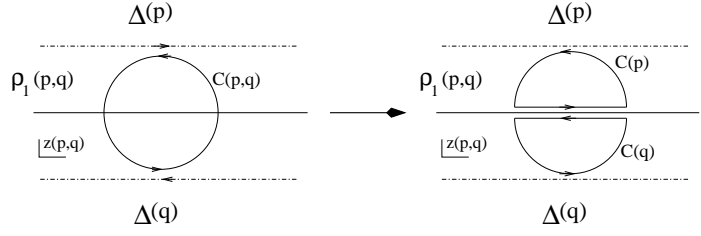}
  \caption{A small integration contour intersecting the $(p,q)$ edge
    of the ribbon graph.}
  \label{fig:int_bound}
\end{figure}

Eq. \eqref{eq:pq_cur} allows to associate a unique copy of the chiral
algebra $\mathcal{W}(p,q)$ to each pairwise adjacent pairs of BCFTs.
To this end, let us now consider a small integration contour crossing
the $(p,q)$ boundary as in figure \ref{fig:int_bound}.

Exploiting the continuity condition along the boundary, the following
holds:
\begin{multline}
  \label{eq:Wnpq}
  \mathbb{W}_n^{(p,q)}  \,=\, \frac{1}{2 \pi i}\oint_{C(p,q)}d z(p,q)\ z(p,q)^{n + 1}
  \mathbb{W}_{(p,q)} (z(p,q)) \,=\,\\
   \frac{1}{2 \pi i}\oint_{C(p)} d z(p,q)\,z(p,q)^{n + 1}
   \mathbf{W}_{\Omega_{A(p)}}
  (z(p,q)) \,+\,\\ \frac{1}{2 \pi i}\oint_{C(q)} d z(p,q) \,z(p,q)^{n +
    1}{\Omega'}^{A(p) B(q)}\mathbf{W}_{\Omega_{B(q)}} (z(p,q)).
\end{multline}

With eqns. \eqref{eq:pq_cur} and \eqref{eq:Wnpq} we have now
introduced all the main ingredients we need in order to coherently
define a full-fledged boundary conformal field theory on the whole
surface $M_\partial$.  As a matter of fact, we can associate to each
$(p,q)$ pair of BCFTs defined on cylindrical ends, which are adjacent
along a ribbon graph edge, a unique Hilbert space of states
$\mathcal{H}^{(p,q)}$; this, can be determined through the action of
chiral modes \eqref{eq:Wnpq} on a true vacuum state, whose existence
is granted per hypothesis. As usual, $\mathcal{H}^{(p,q)}$ gets
decomposed into a direct sum of subspaces $\mathcal{H}_{\lambda(p,q)}$
which are carrier of an irreducible representation of the
$\mathcal{W}(p,q)$ algebra itself. Exploiting the state-to-field
correspondence, we can associate to each highest weight state in
$\mathcal{H}_{\lambda(p,q)}$ a primary field which we shall refer to
as \emph{Boundary Insertion Operator} such that
\begin{equation} 
\label{eq:bio}
\psi_{\lambda(q,p)}^{A(p) B(q)}(x(q,p)) \,=\,
\psi_{\lambda(p,q)}^{B(q) A(p)}(x(p,q)),
\end{equation}
where $x(q,p)\,=\,\Re\left[ z(q,p)\right]$. 
In \eqref{eq:bio} the notation is chosen with the following convention: 
$\lambda(p,q)$ is the representation label, while decoration with indexes
$A(p)$ and $B(q)$ points out that the switch in boundary conditions actually refers to all
parameters which specify the boundary assignation (to quote, in the case
of bosonic string on a toroidal background, it will act both
on the glueing automorphism and on the Wilson line/brane position).

Unfortunately, at this stage, BIOs are only purely formal objects.  At
most, since they are primary operators, we can adopt a description in
terms of Chiral Vertex Operators (CVO); in this framework we can
interpret them as a map from $\mathcal{H}^{(O)}(p) \otimes
\mathcal{H}_{\lambda(p,q)}$, the tensor product between the space of
states on \cyl{p} with prescribed boundary conditions $A(p)$ and the
Fock space of state on the strip into $\mathcal{H}^{(O)}(q)$, the
space of states on \cyl{q} with prescribed boundary conditions $B(q)$.

In order to ``transform'' the functions \eqref{eq:bio} in a concretely
useful tool from a field theoretical perspective, we must analyze in
detail also their analytic and algebraic description.
This will be the guiding idea underlying the remaining discussions in
this manuscript.

As a starting comment we point out that, since BIOs live on the ribbon
graph, their interactions must be guided by the trivalent structure of
$\Gamma$. Hence, it is useful to summarize here a few results of an
exhaustive related analysis on BIOs' correlators \cite{gili-phd},
which shows how the above introduced geometric structure is sufficient
to provide all the fundamental data defining their interaction.

Exploiting the CVO analogue structure, we are able to associate a
well-defined conformal dimension to each element as in \eqref{eq:bio}
which coincides with the highest weight of the $V_\lambda(p,q)$ module
of the Virasoro algebra:
\begin{equation}
  \label{eq:BIOcd}
  H(p,q) \,=\,  \frac{1}{2} \lambda^2(p,q). 
\end{equation}

\begin{figure}[!t]
  \centering
  \includegraphics[width=.9\textwidth]{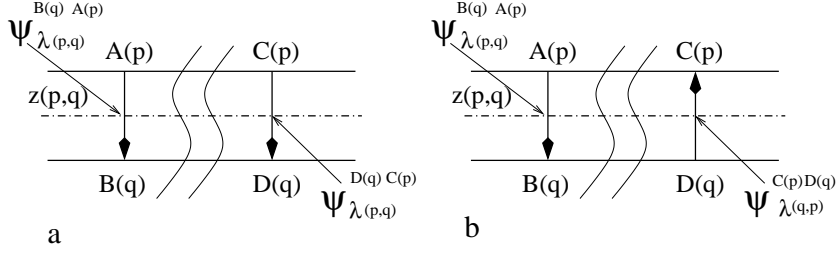}
  \caption{Two-points function of Boundary Insertion Operators.}
  \label{fig:BIO_2points}
\end{figure}

Let us deal with the two-point functions between BIOs. To this end, we
can exploit explicitly the ribbon graph structure which suggests that
they can be introduced as a well-defined concept along any edge
$\rho^1(p,q)$ shared between the cylinders \cyl{p} and \cyl{q}.
Accordingly, we must take into account two different scenarios: in the
first one two operators both mediate a change between boundary
conditions in the ``$p$-to-$q$'' direction, while in the second case,
one mediates in the ``$p$-to-$q$'', the other in the ``$q$-to-$p$''
(see figure \ref{fig:BIO_2points}).

Out of \eqref{eq:bio} and out of conformal invariance, both kind of
correlators have the same analytic form:

\begin{multline}
\label{eq:2points}
  \langle 
  \psi_{\lambda(p,q)}^{B(q) A(p)} (x_1(p,q)) 
  \psi_{\lambda'(q,p)}^{C(p) D(q)} (x_2(q,p))
  \rangle \,=\, \\
  \langle 
  \psi_{\lambda(p,q)}^{B(q) A(p)} (x_1(p,q)) 
  \psi_{\lambda'(q,p)}^{C(p) D(q)} (x_2(q,p))
  \rangle \,= \\
  \frac{b_{\lambda(p,q)}^{B(q) A(p)}
    \delta_{\lambda(p,q) \lambda'(p,q)}
    \delta^{A(p) C(p)}
    \delta^{B(q) D(q)}}
  {|x_1(p,q)\,-\,x_2(p,q)|^{2H(p,q)}},
\end{multline}
where $H(p,q)$ satisfies the identity \eqref{eq:BIOcd} and where each
$b_{\lambda(p,q)}^{B(q) A(p)}$ is a real normalization factor.

\begin{figure}[!t]
  \centering
  \includegraphics[width=\textwidth]{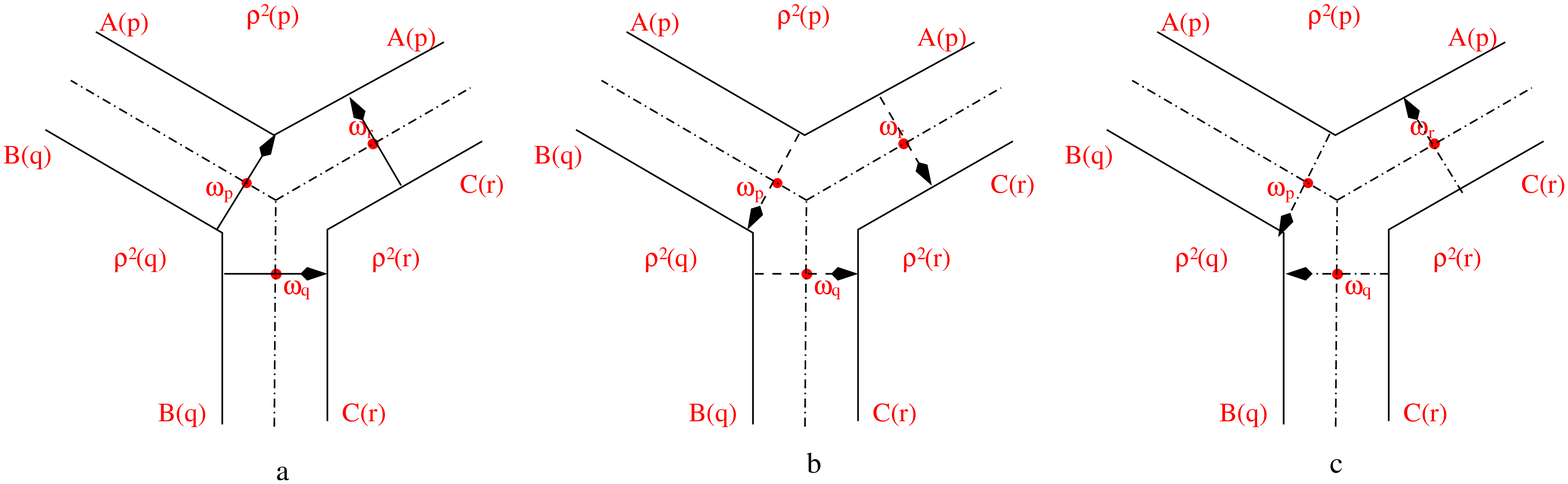}
  \caption{Operator Product Expansion between Boundary Insertion
  Operators.}
  \label{fig:BIO_ope}
\end{figure}

Switching now to the three-points function, its structure is mainly
driven by the operator product expansion calculated in the hypotheses
that BIOs are inserted near any but fixed of the $N_2$ trivalent
vertexes of the ribbon graph (see fig. \ref{fig:BIO_ope}). Thus, let
us take three points in an infinitesimal open neighborhood with radius
$\epsilon$ of a vertex $\rho^1(p,q,r)$, chosen as the origin of a
suitable local chart \cite{Carfora1}.  Furthermore let us denote their
coordinates as $\omega_r$, $\omega_p$ and $\omega_q$ and let us focus
our attention on three fields $\psi_{\lambda(r,p)}^{A(p) C(r)}
(\omega_r)$, $\psi_{\lambda(p,q)}^{B(q) A(p)} (\omega_p)$ and
$\psi_{\lambda(q,r)}^{C(r) B(q)} (\omega_q)$ which mediate pairwise
the boundary conditions respectively between $\partial\rho^2(r)$ and
$\partial\rho^2(p)$, $\partial\rho^2(p)$ and $\partial\rho^2(q)$,
$\partial\rho^2(q)$ and $\partial\rho^2(r)$ (as usual, the direction
of
the action of BIOs is implicitly encoded in the notation).\\
In the limit $\epsilon \rightarrow 0$ the product of the two fields
$\psi_{\lambda(r,p)}^{A(p) C(r)}(\omega_r)$ and
$\psi_{\lambda(q,r)}^{C(r) B(q)}(\omega_q)$ will mediate the change in
boundary conditions from $B(q)$ to $A(p)$. Thus the OPE of these two
fields must be expressed as a function of a $\psi_{\lambda(q,p)}^{A(p)
  B(q)}$-type field:
\begin{multline}
  \label{eq:ope1}
  \psi_{\lambda(r,p)}^{A(p) C(r)}(\omega_r) 
  \psi_{\lambda'(q,r)}^{C(r) B(q)}(\omega_q)
  \,\sim\, \\
  \sum_{\lambda''(q,p) \in \mathcal{Y}}
  \mathcal{C}^{A(p) C(r) B(q)}_{\lambda(r,p) \lambda'(q,r) \lambda''(q,p)}
  |\omega_r\,-\,\omega_q|^{H(q,p)\,-\,H(r,p)\,-\,H(q,r)}
  \psi_{\lambda''(q,p)}^{A(p) B(q)} (\omega_q),
\end{multline}
being $\mathcal{C}^{A(p) C(r) B(q)}_{\lambda(r,p)\lambda'(q,r)
\lambda''(q,p)}$ the OPE coefficients. 
This overall scenario is depicted by the continuous arrows in fig.
\ref{fig:BIO_ope}. The same holds in all other cases.

Hence the complete description of the interaction among the $N_0$
different BCFTs requires the determination of an explicit expression
for $\mathcal{C}^{A(\cdot) B(\cdot) C(\cdot)} _{\lambda(\cdot,\cdot)
  \lambda'(\cdot,\cdot) \lambda''(\cdot,\cdot)}$ with
$A,\,B,\,C\,\in\,\mathcal{A}$ and
$\lambda,\,\lambda',\,\lambda''\,\in\,\mathcal{Y}$. Luckily enough this
task is partly tractable. As a matter of fact, one can show that both
the OPE coefficients and $b_{\lambda(p,q)}^{B(q) A(p)}$ satisfy a set
of cyclic properties and sewing constraints that, thanks to the
trivalent structure and the variable connectivity of $\Gamma$, have an
high resemblance with similar problems in ordinary BCFT (see
\cite{Lewellen:1991tb}). In particular, in section \ref{sec:ratbcft}
we will show that, through a suitable choice of toroidal background,
it is possible to exploit the variable connectivity of $\Gamma$ and
BIO four-point functions to fix the algebraic form of the above data.

\section{Open String Gauge Theory on a RRT}
\label{sec:osgauge}
An enhancement of our model, which could also play a pivotal role in
gauge/gravity correspondences, calls for the inclusion of open string
gauge degrees of freedom (propagating) along the boundaries of
$M_\partial$.

To this avail, let us follow usual techniques in open string theory
where a non Abelian gauge theory can be naturally included into an
open string model by means of a suitable assignation of non-Abelian
Chan Paton factors at the open string endpoints.

Thus, let us decorate each \cyl{p} with  suitable $U(N)$
Chan-Paton factors (let us remember that $M_\partial$ is oriented). 
The full  string states now transform under the $N \otimes \overline{N}$
representation, namely the adjoint of $U(N)$. Consequently the generators
 $T^a$, $a = 1, \ldots, \text{dim}\left[\mathfrak{u}(N)\right] = N^2$ 
label the string states now belonging to the tensor product between the 
Fock space associated to the BCFT on the cylinder and the carrier space 
of the $N \times \overline{N}$ representation {\it i.e.} a direct sum of 
the subspaces\footnote{The
reader should bear in mind that $\mathcal{H}_{\lambda}$ is still the
sub-Hilbert space first appeared in \eqref{eq:chirdec}.} 
$\mathcal{H}_{\lambda}\otimes N \otimes \overline{N}$
constructed out of the ground state $\vert 0, \lambda ; i \bar{j} 
\rangle$. Conversely we will refer as $(T^a)^i_{\bar{j}}$ to the matrix
elements which specify the charges $q^i$ and $q_{\overline{j}}$
created at strings/cylinder endpoints. 

In the $k$-th sector - $k = 1, \ldots, N_0$ - the net effect of a
dynamical background gauge field $A_\alpha$ is accounted for including
in the Polyakov path integral, for each boundary component, a Wilson
line term $Tr\left[P\,exp(-S_A)\right]$, where $S_A$ represents the
following boundary condensate:
\begin{equation}
\label{eq:phot_bc}
S_A \,=\,  \int d \tau\, A_\alpha\, \partial_\tau X^\alpha.
\end{equation}

Exploiting conformal invariance, the associated $\beta$-functions
vanish and, in particular, the equation $\beta_A = 0$ reduces, at the
leading order in the $\sigma$-model expansion, to the Yang-Mills
equation \cite{Johnson:Dbranes}.

The inclusion of gauge degrees of freedom forces us to slightly modify
the overall picture on the interaction along the ribbon graph for the
BCFTs defined on adjacent cylinders. Since the latter, glued along
one edge of the ribbon graph, have opposite orientation, the
associated kinematical degrees of freedom must fall into opposite
representations of the gauge group and, hence, the whole graph acquires
a well defined gauge coloring mirroring that of $M_\partial$'s boundary.

Concerning the Fock space for a conformal theory on a shared edge, we
can proceed as in the previous section. However we must take into
account that, due to the components in the $N\otimes\bar N$ space, the
states rotate with the action of the adjoint representation of the
gauge group, a fundamental datum to take into account whenever we deal
with the limit where such states are interpreted as particles. Hence
we seek for an object out of the tensor product between the two
original Chan-Paton factors thought as elements in the gauge algebra.
The only products of this kind are the symmetric and antisymmetric
ones between the generators \footnote{Unless stated otherwise, we
  adopt the following conventions for the algebra structure constants:
$$f^{abc}=\frac{2}{i}Tr\left(\left[T^a,T^b\right]T^c\right),\quad
d^{abc}=2Tr\left(\left\{T^a,T^b\right\}T^c\right).$$}:
\begin{equation*}
  T^a_{i l}(p,q) \,=\,
  \sum\limits_{b,c=1}^{N^2}\frac{i}{2}f^{abc}\,\left[T^b_{ij}(p)\,,\,T^c_{jl}(q)\right]
  \,+\,\sum\limits_{b,c=1}^{N^2}\frac{d^{abc}}{2}\left\{T^b_{ij}(p)\,,
  \,T^c_{jl}(q)\right\}.
\end{equation*}

Accordingly each BIO belonging to the $\rho^1(p,q)$ BCFT spectrum must 
be decorated by a $\mathfrak{u}(N)$ generator $T^a_{i l}$.
Hence, denoting the conformal structure of BIOs in formula
\eqref{eq:bio} with a collective subscript $\Xi(p,q)$, the new 
non-Abelian BIOs will be matrix-valued functions 
$\psi_{\Xi(p,q)}^a\doteq T^a \psi_{\Xi(p,q)}$. 
As a first manifest consequence of these remarks, correlation functions 
between BIOs acquire a prefactor, namely the trace of the relevant gauge 
algebra generators.

The first big difference from the analysis in the previous section lies
instead in a ``simplification'' of the two-points function since only the 
product between two fields mediating along opposite directions is 
meaningful, hence halving the possible cases. 

To summarize, the modified BIOs algebra can be recast (anti)symmetrizing 
the product of generators:
\begin{multline}
\label{eq:abope}
  \psi^a_{\Xi_1(r,p)}(\omega_r) \,
  \psi^b_{\Xi_2(q,r)}(\omega_q)
  \,\sim\,
  \frac{1}{2}\sum_{\Xi_3}\sum\limits_{c=1}^{N^2}
  \left\vert
    \omega_r\,-\,\omega_q
  \right\vert^{H(q,p)\,-\,H(r,p)\,-\,H(q,r)} \times \\
  \left(
    i f^{a b c} + d^{a b c}
  \right)
  \mathcal{C}(\Xi_1(r,p),\Xi_2(q,r),\Xi_3(q,p))
  \psi^c_{\Xi_3(q,p)}(\omega_q),
\end{multline}
being $\mathcal{C}(\Xi_1(r,p),\Xi_2(q,r),\Xi_3(q,p))$ the previously
introduced operator algebra fusion coefficients here written with the
novel multi-index notation.

\subsection{Coupling with background gauge potential}
Our next aim is to analyze the new kinematical background emerging
after the inclusion of gauge degrees of freedom. Within this respect we 
will show that the natural coupling with background gauge fields can be
rephrased as a move between different orbits in the moduli space of
toroidal compactifications. This will allow us to provide a complete
characterization of BIOs dynamic specifying the coefficients
$\mathcal{C}(\Xi_1(r,p),\Xi_2(p,q),\Xi_3(q,r)) $ in \eqref{eq:abope}.

Thus, let us consider a $D$-dimensional background in which each
direction $X^\alpha$, $\alpha=0,\,\ldots,\,D-1$ is compactified on a
circle of radius $\frac{R^\alpha(p)}{l(p)}$.  Consistently with the
gluing process, let us assume the injection maps 
obey to arbitrary boundary conditions on the inner
boundary of \cyl{p} (in the annuli picture).  On the outer
one, let us assume to have $n+1$ directions satisfying Neumann
boundary conditions and $D - n - 1$ directions obeying Dirichlet
ones:
\begin{equation*}
\left\{ X^\alpha \right\}
\,\doteq\,
\left\{ X^i ,\, X^m\right\}\quad
\text{with} \;
i\,=\,0,\,\ldots,\,n,\,\, 
m\,=\,n + 1,\,\ldots,\,D - 1. 
\end{equation*}

To rephrase in a stringy language, we are dealing with a
D$n$-brane lying along the $X^0,\,\ldots,\,X^n$ directions assumed to be
coincident with the world-volume parameters
$\xi^0,\,\ldots,\,\xi^n$ {\it i.e.} $\xi^i = X^i$ for all
$i=0,\ldots,n$.
 
To endow the brane with an interesting dynamic, we have to couple the
model to a background gauge field living on its worldvolume: this can
be worked out introducing the following boundary action
\cite{Callan:1988wz, Callan:1995xx}:
\begin{equation}
  \label{eq:gaugeba}
  S_A\,=\, \int\limits_0^T d \tau 
  \left[
    \sum_{i=0}^{n} A_i(\vec{X}) \partial_\tau X^i  
    \,+\,
    \sum_{m = n + 1}^{D - 1} \phi_m(\vec{X}) \partial_\sigma X^m  
  \right],
\end{equation}
where $T$ is an unspecified (and, at this stage, irrelevant) finite
real number, where we have chosen the boundary to lay at constant
$\sigma$ and where $\vec{X}=\left\{X^0,...,X^n\right\}$.  The
$A_i(\vec{X})=\sum\limits_{a=1}^{N^2} A_i^a(\vec{X}) T^a $ are Lie
algebra valued gauge fields on the D-brane, while the entries of the
$N \times N$ matrices $\phi_m(\vec{X})=\sum \limits_{a=1}^{N^2}
\phi_m^a(X^i) T^a$ are real scalars from the world volume point of
view; the latter describes the motion of the brane in the transverse
space.

For the sake of simplicity, let us now assume the brane static in the
transverse space {\it i.e.} $\phi_m = \mathbf{0}_{N \times N}$ $\forall
m = n+1 \ldots D-1$.  Moreover let us take constant electric and
magnetic fields along the brane worldvolume. Accordingly, the boundary 
term reads:
\begin{equation}
  \label{eq:gaugebaF}
  S_A=\sum\limits_{i,j=0}^n F_{i j}
  \int\limits_0^T d \tau X^j \partial_\tau X^i,  
\end{equation}
being $F_{ij}$ the constant field strength out of $A_i(\vec{X})$.

For further convenience, let us specialize to the Abelian subsector. Such specific case can 
be achieved including in each Neumann direction of the T-dual theory a
Wilson line such that $A_i(\vec{X})$ is at the same time a pure gauge and a
diagonal matrix {\it i.e.} $U(N)$ symmetry is broken into $U(1)^N$. At 
spacetime level, the global effect will be a
displacement of the position of  N D-branes which, accordingly, 
entails us to deal only with N separated D-branes.

At a Lagrangian level, on each $(p)$-subsector the above reasoning
translates in the coupling between the open 
string with a different electromagnetic potential $A_i(p; \vec{X})$; 
hence  \eqref{eq:gaugebaF} is equivalent to:
\begin{equation}
  \label{eq:gaugebaB}
  S_{A(p)}\,=\, 
  \sum\limits_{i,j=0}^n F_{i j}(p)
  \int\limits_{\Delta^*_{\epsilon(p)}} d \zeta(p) d \bar{\zeta}(p)
  \partial X^i(p) \, 
  \bar{\partial}\, \overline{X}^j(p).
\end{equation}
Comparing last formula with \eqref{eq:action}, we can state
that, in the Abelian subsector, the net effect of \eqref{eq:gaugebaB}
is to move to a different point in the flat toroidal
background moduli space \cite{yegulalp,Green:1995ga}:
\begin{center}
\begin{tabular}{|ccc|}
\hline
Config. $\mathcal{A}$ & & Config. $\mathcal{B}$\\
\hline
$G_{\alpha \beta} \,=\, \mathbb{I}_{D \times D}$ & & 
$G_{\alpha \beta} \,=\, \mathbb{I}_{D \times D}$ \\
$B_{\alpha \beta} \,=\, 0 \: \forall \alpha,\,\beta$ 
& $\Longleftrightarrow$ &
$B_{\alpha \beta} \,=\, 4 \pi \Lambda_{\alpha \beta}$ \\
$F_{\alpha \beta} \,=\, \Lambda_{\alpha \beta}$ 
& &
$F_{\alpha \beta} \,=\, 0 \: \forall \alpha,\,\beta$\\
\hline
\end{tabular}
\end{center}

On this wise, the description of the new kinematical background
directly resides in the choice of a particular point in the moduli
space of inequivalent toroidal compactifications in a $D$-dimensional
space, with associated suitable values of the background matrix $E$
entries (see formula \eqref{eq:bm}), namely
\cite{Johnson:Dbranes,Giveon:1994fu}
\begin{equation}
  \label{lattice_modspace}
  \mathcal{M}
  \,=\, 
  O(D,D,\mathbb{Z}) \backslash O(D,D) /[O(D) \times O(D)].
\end{equation}

The different orbits in $\mathcal{M}$ give rise to different
theories in which the fundamental $U(1)_L \times U(1)_R$ current
symmetry can be enhanced to different symmetry groups of rank at least
$D$ playing the role of gauge group in the target space.

Thus, higher dimensional toroidal compactifications are described by
non-trivial background fields $B$ and $G$ and, in such a given
background, the maximally enhanced symmetry points are those fixed
under the action of $O(D,D,\mathbb{Z})$\footnote{
$O(D,D,\mathbb{Z})$ is the generalized T-duality group and its action on
a background matrix $E$ can be represented in terms of an element 
$M \in SL(D, \mathbb{Z})$ and  of an antisymmetric integer valued 
matrix $\Theta$:
\begin{equation}\label{eq:rote}
E'\,=\, M^t \,(E + \Theta)\, M.
\end{equation}}.
In these special points in which \eqref{eq:rote} provides $E' = E$, we 
can represent extended target space symmetries with respect to a semisimple 
simply laced Lie algebra of total rank $D$.

In particular, the maximally enhanced symmetry background can be
chosen in the following way \cite{Giveon:1994fu}: if
$C_{\alpha\beta},\,\alpha, \beta = 1, \ldots, D$, stands for the
Cartan matrix of the semisimple simply laced Lie algebra of total rank D, then
we must fix:
\begin{subequations}
\label{bkBG}
  \begin{gather}
\label{bkBG1}
    G_{\alpha \beta} = \frac{1}{2}  C_{\alpha \beta}  \\
    B_{\alpha \beta} = G_{\alpha \beta} \,\,\forall\, \alpha \,>\,
    \beta,
    \quad
    B_{\alpha \beta} = -\, G_{\alpha \beta} \,\, \forall\, \alpha \,<\, 
    \beta,    \quad
    B_{\alpha \alpha} = 0,\label{bkBG2}
  \end{gather}
\end{subequations}
hence the background matrix $E = G + B\in SL(D,\,\mathbb{Z})$ and
it is fixed under the action \eqref{eq:rote} of $O(D,D,\mathbb{Z})$. 

Still in this framework though in a more specific example, let us 
consider an $O(D,D,\mathbb{Z})$ transformation acting by $M = E^{-1}$ and 
$\Theta = E^\dag + E$. Hence $E' \,=\, E^{-1}$ and, whenever $G =
\mathbb{I}_{D \times D}$ and $B = \mathbf{0}_{D \times D}$, this is
exactly the case $(SU(2)_L \times SU(2)_R)^D$. 
 
Accordingly, the extended symmetry
group associated with the boundary action \eqref{eq:gaugebaB} will be:
\begin{equation}
\label{eq:sgrup}
\mathbb{G}_D \,=\, ({G}_{p+1} \,\times{G}_{p+1}) \,\times\,  (SU(2)_L \times
  SU(2)_R)^{D - p - 1}. 
\end{equation}

\section{Redefinition of the BCFT in the rational limit}
\label{sec:ratbcft}

The lesson we can draw from the previous section is the existence of
an equivalence between the description of the interaction for each
open string with a gauge field living on the brane world-volume and
the choice of a point in the moduli space of toroidal
compactifications
characterized by a non vanishing value of the Kalb-Ramond field.\\
Moreover we have shown that, among all such choices, we can pick up in
$\mathcal{M}_d$ determined as in \eqref{lattice_modspace} some special
points fixed under the action of the generalized T-duality group
$O(D,D,\mathbb{Z})$. In these, the emergence of the extended symmetry
group \eqref{eq:sgrup} is an hint of the equivalence at a quantum
level between such a theory of compactified $D$-free scalar bosons and
the Wess-Zumino-Witten model associated to the level $k=1$ simply
laced affine algebra $\hat{\mathfrak{g}}$, {\it i.e.} the affine
extension of the algebra $\mathfrak{g}$ characterized by the Cartan
matrix entries \eqref{bkBG}.

In more detail, whenever $E^\prime=E$ out of \eqref{eq:rote}, the center
of mass string momentum is such that the set of chiral currents of the
associated bulk conformal field theory gets enlarged. The new set of
arising chiral fields together with the old ones provides for the 
closure of the $\hat{\mathfrak{g}}$ affine algebra. Moreover, since the
Virasoro algebra belongs to the double covering of the chiral one, it
is possible to reorganize the infinite sets of highest weights
representations into finitely many chiral algebra ones, in particular
those appearing when the level $k$ is fixed to 1.

Thus, within this specific choice for the background matrix, in order
to analyze what it is the overall behavior of our model with $n+1$
Neumann and $D-n-1$ Dirichlet directions, let us choose a $G_r\times
G_r$ factor in \eqref{eq:sgrup} describing the enhanced symmetry group
of a generic but fixed set of $r$ compact directions. According to the
previous remark $G_r$ is nothing but the universal covering group
generated by exponentiation from the rank-$r$ Lie algebra
$\mathfrak{g}$, and this model is equivalent, at a quantum level, to
the $\hat{\mathfrak{g}}_{1}$-WZW model.

We will show that this choice allows us to introduce a specific
parametrization of boundary conditions which, through a careful
analysis, completely characterize the action of the glueing
automorphism introduced in \eqref{eq:ifr}, hence also the action of
Boundary Insertion Operators.

Before getting into the detail of the analysis we shall address in this
section, let us fix a few notations and conventions. We denote with $P_+
^k(\hat{\mathfrak{g}})$ the set of all finitely many integrable level-$k$ 
highest weight representations of $\hat{\mathfrak{g}}_k$.  The associated 
highest weights can be characterized by their Dynkin labels 
(non-negative integers)
$\hat\lambda\doteq[\lambda_0; \lambda_1, \ldots, \lambda_r] = 
[\lambda_0, \lambda]$,
\ie{}\footnote{In this notation $\lambda$ denotes the finite part of the 
weight \ie{} it is an
integrable highest weight of the parent finite algebra $\mathfrak{g}$.
As a side effect let us pinpoint that it does not keep track of 
the $- L_0$ operator eigenvalue.} their expansion coefficients in the 
basis of the fundamental weights $\hat{\omega}_l$, $l = 0, \ldots, r$. 

Representations of $\hat{\mathfrak{g}}_k\in P_+^k(\hat{\mathfrak{g}})$ 
are those satisfying the
constraint $k \geq (\hat\lambda,\theta)$, where $\theta$ is the highest
root of $\mathfrak{g}$, while $(\cdot,\cdot)$ is the scalar
product naturally induced by its Killing form.

Furthermore we refer to $\chi_{\hat{\lambda}}$ - $\hat{\lambda}\in
P_+^k\left(\hat{\mathfrak{g}}\right)$ - as the characters which carry
a representation of the modular group whose properties are partially
encoded in $\mathcal{S}^{ext}$:
\begin{equation*}
  \chi_{\hat{\lambda}}(- \frac{1}{\tau}) \,=\, \sum_{\hat{\mu} \in
  P_+^k} \mathcal{S}^{ext}_{\hat{\lambda} \hat{\mu}} \,
  \chi_{\hat{\mu}}(\tau). 
\end{equation*}

Let us now focus on a specific scenario we are interested in, namely 
$k = 1$. The only highest weight representations entering in
$P_+^1(\hat{\mathfrak{g}})$ are those generated by the highest weights
$\hat{\omega}_I$ whose correspondent simple root $\hat{\alpha}_I$ has
unit comark. Since the one generated by the basic fundamental weight 
$\hat{\omega}_0$ always belongs to $P_+^1(\hat{\mathfrak{g}})$, let us 
rewrite the set of its elements as
\begin{equation*}
  P_+^1(\hat{\mathfrak{g}}) \,=\, \{ \hat{\omega}_I \} 
\,=\, \{ \hat{\omega}_0, \hat{\omega}_i \}.
\end{equation*}

The explicit set of $\hat{\omega}_I \in P_+^1(\hat{\mathfrak{g}})$ for
$\hat{\mathfrak{g}}$ being a simply laced algebra is reported for later
convenience and for sake of completeness in table \ref{tab:ADE1weights}. 

\begin{table}[!t]
  \centering
  \begin{tabular}{|c|l|c|c|}
\hline
$\hat{\mathfrak{g}}$& $\hat{\omega}_I \in P_+^1(\hat{\mathfrak{g}})$
& $B(G) \sim \mathcal{O}(\hat{\mathfrak{g}})$ & $h^\vee$\\
\hline
$\hat{A}_r$ & $\hat{\omega}_0,\, \hat{\omega}_1,\, \ldots,\,
\hat{\omega}_r$ & $\mathbb{Z}_{r+1}$ & $r + 1$\\
$\hat{D}_{r = 2l}$ & 
$\hat{\omega}_0,\, \hat{\omega}_1,\, 
\hat{\omega}_{r-1},\, \hat{\omega}_r$ &
$\mathbb{Z}_{2}\times\mathbb{Z}_{2}$ & $2 r - 2$ \\
$\hat{D}_{r = 2l + 1}$ & 
$\hat{\omega}_0,\, \hat{\omega}_1,\, 
\hat{\omega}_{r-1},\, \hat{\omega}_r$ &
$\mathbb{Z}_{4}$ & $2 r - 2$ \\
$\hat{E}_6$ &  $\hat{\omega}_0,\, \hat{\omega}_1,\, \hat{\omega}_5$
& $\mathbb{Z}_{3}$ & 12 \\
$\hat{E}_7$ &  $\hat{\omega}_0,\, \hat{\omega}_6$ &
$\mathbb{Z}_{1}$ & 18\\
$\hat{E}_8$ &  $\hat{\omega}_0$ & $\mathbb{I}$ & 30\\
\hline
  \end{tabular}
  \caption{This table reports fundamental weights belonging to
    $P_+^1(\hat{\mathfrak{g}})$, the outer automorphism group
    $\mathcal{O}(\hat{\mathfrak{g}})$, and the dual Coxeter number for
    $\hat{\mathfrak{g}}$ being a 
    simple laced affine untwisted Lie algebra.} 
  \label{tab:ADE1weights}
\end{table}

Within this framework, the bulk theory can then be fully characterized
by all the properties of $\hat{\mathfrak{g}}_1$-WZW model. The infinite
series of holomorphic and antiholomorphic Verma modules can be
reorganized to write the Fock space of the parent bulk theory as the
direct sum of the finitely many moduli of the affine Lie algebra:
\begin{equation}
  \label{eq:gWZW_Hspace}
  \mathcal{H}^{(C)} \,=\,
  \bigoplus_{\hat{\omega}_I \in P_+^1(\hat{\mathfrak{g}})}
  \mathcal{H}_{\hat{\omega}_I}^{\hat{\mathfrak{g}}_1}
  \,\otimes\,
  \overline{\mathcal{H}}_{\hat{\omega}_I}^{\hat{\mathfrak{g}}_1},
\end{equation}
being $\mathcal{H}_{\hat{\omega}_I}^{\hat{\mathfrak{g}}_1}$ the
subHilbert space associated with $\hat{\omega}_I$.

We will denote (the holomorphic part of) the primary
fields associated to the highest weight state in
$\mathcal{H}_{\hat{\omega}_I}^{\hat{\mathfrak{g}}_1}$
with\footnote{From this stage on, we shall trade the subscript
  $\hat{\omega}_I$ in the operators with $\hat{I}$ in order,
  hopefully, to provide a simpler notation.}:
\begin{equation*}
\phi_{\hat{I}(p)}(\zeta(p)).
\end{equation*}
Their components $\phi_{[\hat{I}(p),\,m]}(\zeta(p))$, $m \,=\,
1,\,\ldots,\,\text{dim}\hat{\omega}_I$, fill the level-0 (in sense of
$L_0$ eigenvalue) subspace of
$\mathcal{H}_{\hat{\omega}_I}^{\hat{\mathfrak{g}}_1}$, which we will
denote with $\mathcal{V}_{\hat{\omega}_I}^0$. These last subspaces carry
an irreducible representation of the horizontal subalgebra of
$\hat{\mathfrak{g}}_1$:
\begin{equation}\label{eq:cla5}
  \mathbf{X}^{\hat I}_{J_0}\,:\; \mathcal{V}_{\hat{\omega}_I}^0
  \,\rightarrow\, \mathcal{V}_{\hat{\omega}_I}^0,
\end{equation}
being $J_0$ a generic element of $\mathfrak{g}$ \cite{Recknagel:1998ih}.

Since we are ultimately dealing with rational conformal field theories 
associated to WZW models, we shall exploit their similarity with conformal
minimal models. Thus, to extend them on a surface
with boundary, we can adopt Cardy's construction: a set of boundary
conditions that we can consistently define on the boundaries are labelled
exactly by the modules of the chiral algebra entering into the
Hilbert space. The correspondent boundary states are
\cite{Cardy1}:
\begin{equation}
  \label{eq:cardy_bs}
  \vert\vert \hat{\omega}_I(p)\Rangle \,=
  \sum_{\hat{\omega}_J\,\in\,P_+^1(\hat{\mathfrak{g}})}
  \frac{\mathcal{S}_{\hat{I}\,\hat{J}}}
  {\sqrt{\mathcal{S}_{\hat{0}\,\hat{J}}}} 
  |\hat{\omega}_J(p)\Rangle.
\end{equation}
They obey the glueing condition:
\begin{equation}
  \label{eq:unp_gc}
  \left(
J^a_n \,+\, \overline{J}^a_{-n}
\right)
 \vert\vert \hat{\omega}_I(p)\Rangle \,=\, 0 \qquad \quad \forall\;
 \hat{\omega}_I \,\in\,P_+^1\left(\hat{\mathfrak{g}}\right) 
\end{equation}

However, Cardy boundary states are not sufficient to describe the
plethora of boundary assignations we can coherently fix for a WZW
model with a prescribed bulk action.  A comprehensive description
calls into play deformations techniques of a BCFT
\cite{Recknagel:1998ih}.  As a matter of fact, when we deal with
special points in the toroidal compactifications moduli space, the
presence of the enhanced affine symmetry coincides with the presence
of new massless \emph{open} string states which can be used to deform
the boundary conformal field theory on \cyl{p} \cite{Green:1995ga}. In
particular, if we pick up among these the chiral deformations, \ie{}
induced by chiral operators, these deformations are truly marginal
(for a brief description of key concepts and techniques in BCFT
deformations see \cite{gili-phd}), hence the deformed model will
change from the undeformed one only for a redefinition of boundary
conditions. In this way, starting from an unperturbed Lagrangian, we
are able to describe the full set of boundary conditions we can
adopt\cite{Recknagel:1998ih} by means of an its suitable deformation.

To provide a detailed description, let us represent the closed affine 
algebra generators in terms of the boson fields via the Frenkel-Kac-Segal
construction of the Weyl-Cartan basis of ${\hat{\mathfrak{g}}}_{k=1}$.
In the closed string channel, the left moving and right moving
currents $J^a(\zeta)$, $\bar{J}^a(\bar{\zeta})$ are respectively defined
out of following components:
\begin{subequations}
  \label{eq:cD}
  \begin{gather}
    \label{eq:lcD}
    H^i(\zeta) \,=\, 
    \partial\,X^i(\zeta), \qquad 
    E^\alpha (\zeta) \,=\, c(\alpha) \,
    \ordprod{e^{i\sum\limits_{i=1}^r \alpha_i X^i(\zeta)}},\\
    \label{eq:rcD}
    \overline{H}^i(\bgz) \,=\, 
    \sum\limits_{j=1}^r M^i_j
    \bar{\partial}\,\overline{X}^j(\bgz), \qquad 
    \overline{E}^\alpha (\bgz) \,=\, - \bar{c}(\alpha) \,
    \ordprod{e^{i \sum\limits_{i, j=1}^r\alpha_i M^i_j \overline{X}^j(\bgz)}}.
  \end{gather}
\end{subequations}
 
Here $H^i(\zeta)$ and $\overline{H}^i(\bgz)$ denote the elements in
the maximal torus of the two copies of the chiral algebra
${\hat{\mathfrak{g}}}_{k=1}$, while $\{\alpha\}$ is the set of roots
(positives plus negatives) of the parent semi-simple simply laced Lie
algebra.  The functions $c(\alpha)$ and $\bar{c}(\alpha)$ are
$\mathbb{Z}_2$-valued cocycles; these are operators acting on the Fock
spaces and they depend only upon the momentum part of the free-boson
zero modes. Their inclusion leads the product of the above
currents to satisfy the correct OPE \cite{DiFrancesco}.

In this framework, the vertex operators associated to the new open
string scalar states can be written, in the closed string channel, as
\begin{equation}
\label{eq:openvert}
  S^a_\lambda (u(p)) e^{i \sum\limits_{i=1}^r\lambda_i X^i(\zeta)} \,
  \doteq\, \frac{1}{2}   \bigl.
  \left[ 
    J^a(\zeta) + \overline{J}^a(\bgz)
  \right]\left.
  e^{\sum\limits_{i=1}^r\lambda_i X^i(\zeta)}
  \right|_{|\zeta| = \frac{2\pi}{2\pi - \varepsilon(p)}},
\end{equation}
where, $u(p) = \Re\left[{\frac{2\pi i}{L(p)} \ln[\zeta(p)]}\right]$ is
the coordinate parametrizing the inner boundary of \cyl{p}.

To simplify the notation, from now on, any function dependent upon the
coordinate $u(p)$ implicitly refers to the restriction of an holomorphic 
(or antiholomorphic) map to the locus $|\zeta|=\frac{2\pi}{2\pi - 
\varepsilon(p)}$.

The occurrence of extra massless open string states in equation
\eqref{eq:openvert} indicates the enlargement of the chiral algebra of the
boundary theory. The associated currents
$\mathbf{J}^a(\zeta)$ (which correspond to the vertex operators in
\eqref{eq:openvert} built on the vacuum representation and 
generically defined as in equation \eqref{eq:1al} out of the
holomorphic and antiholomorphic currents $J^a(\zeta)$ and
$\overline{J}^a(\bgz)$), can deform the original theory
with a suitable boundary term $ S_B \,=\, \int d u(p) \sum_a g_a
\mathbf{J}^a(u(p))$.  If we write the 
currents in the Cartan-Weyl basis, it has been shown in
\cite{yegulalp} that the most general such a term will be:
\begin{equation}
  \label{eq:pa}
  {S'}_{g} \,=\,\int\limits_0^{L(p)} d u(p) 
  \,\left(
    \sum_{\hat{\alpha}} g_{\hat{\alpha}}\,e^{i\sum\limits_{i=1}^r\hat{\alpha}_i\,
    X^i(u(p))} 
\,+\, 
    \sum_{i=1}^r g_i \partial_u X^i(u(p))
  \right),
\end{equation}
where $(g_{\hat{\alpha}},g_i)$ are coupling constants and where the
new vectors $\hat{\alpha}$ are related to the simple Lie algebra
$\mathfrak{g}$ roots by means of the relation:
\begin{equation*}
  \alpha_i \,=\, 
  \sum\limits_{j=1}^r(\delta^i_j + M^i_j) 
  \,\hat{\alpha}^j  
\quad \text{where}\quad M\,=\,\frac{G + B}{G - B}. 
\end{equation*}

Since chiral marginal deformations are truly
marginal \cite{Recknagel:1998ih}, the deformed model will change from
the unperturbed one only for a redefinition of boundary conditions
{\it i.e.} glueing automorphism and boundary states.

The effect of such a perturbation on the boundary state is a
rotation with respect to the left-moving zero modes of the
currents \cite{Green:1995ga}:
\begin{equation*}
  \Vert B \Rangle_g \,=\, e^{i \sum\limits_{\hat\alpha}g_{\hat{\alpha}} 
  E_0^{\hat{\alpha}} \,+\,
  i \sum\limits_i g_i H^i_0} \Vert B \Rangle.
\end{equation*}

Thus, according to the previous formula, 
it is possible to describe the full set of boundary states of our
model through a rotation on a fixed one acting as a ``generator''
which is associated with the free (unperturbed) model
\begin{equation}
\label{eq:bs_schom}
  \Vert g \Rangle
 \,=\,g\,
  \Vert B \Rangle_{(free)}
\qquad \Longrightarrow \qquad 
g\,=\, e^{\sum_a g_a J^a_0}.
\end{equation}
They satisfy the perturbed glueing condition:
\begin{equation*}
\left[
J^a_m \,+\, \gamma_{g} (\bar{J}^a_{-m}) 
\right] \Vert g \Rangle = 0, \qquad g\,=\, e^{\sum\limits_b g_b\bar{J}^b_0}  
\end{equation*}
where $\gamma_{g} (\bar{J}^a_{-m}) = e^{- \sum\limits_b
  g_b\bar{J}^b_0} \bar{J}^a_{-m} e^{\sum\limits_b g_b\bar{J}^b_0}$
and they cover the full moduli space of boundary states. 

To provide a
concrete example, let us consider any but fixed Dirichlet direction.
The fixed point in toroidal compactification moduli space is the
$T$-dual radius value, $\frac{R(p)}{l(p)} = \sqrt{2}$, and, as
explained at the end of the previous section, the one-boson bulk CFT
becomes equivalent to $\hat{\mathfrak{su}}(2)_{1}$-WZW model. In this
case, the free theory has associated Neumann boundary condition with
null Wilson line parameter $\tilde{t}_-(p)=0$, {\it i.e.} the
free-theory boundary state is $\Vert N(0) \Rangle_{s.d.}$. Hence, the
Dirichlet boundary state is obtained exploiting the particular choice
of the perturbing boundary action whose associated $SU(2)$ element is
$g = e^{- i \pi J_0^1}$:
\begin{equation}
  \label{eq:dir_bs}
  \Vert D(0) \Rangle_{s.d.} \,=\, e^{- i \pi J_0^1} \Vert N(0) 
  \Rangle_{s.d.}.
\end{equation}

Going back to the general case of $r$ directions described through the
affine algebra $\hat{\mathfrak{g}}_1$, the boundary action in
\eqref{eq:pa} perturbs the spectrum of boundary operators of each
independent conformal theory defined on a single cylindrical end.  In
this connection, the rotation of a boundary field $\psi_i$ induced on
a boundary operator $\psi_{\hat{I}(p)}$ (since we are not moving onto
a definite representation, we can omit the quantum number $m$) by a
boundary term like that in equation \eqref{eq:pa} is
\cite{Recknagel:1998ih}:
\begin{gather}\notag
  \tilde{\psi}_{\hat{I}(p)}(u(p))\,=\,
  \left[
    e^{\frac{1}{2} J} \psi_{\hat{I}(p)}
  \right](u(p))  \,\doteq\\
  \label{eq:boh} 
  \doteq\sum_{n = 0}^\infty
  \frac{\lambda^n}{2^n\,n!}
  \oint_{C_1}\frac{d v_1}{2\pi} 
  \cdots
  \oint_{C_n}\frac{d v_n}{2\pi}\,
  \psi_i (u)\,J(v_1)\,\cdots\,J(v_n), 
\end{gather}
where each $C_l$ is a small circle surrounding the $J$-insertion
points.
Let us now think at the previous expression as suitably inserted into 
bulk and boundary fields correlators. Hence we can compute explicitly the 
expression of $\tilde{\psi}_{\hat{I}(p)}$ thanks to the self locality of 
boundary operators and to the OPE between the truly marginal fields in 
the chiral algebra and a boundary operator :
\begin{equation*}
  \mathbf{J}(u')\psi_{\hat{I}}(u) \sim
\frac{\mathbf{X}_{J_0}^{\hat{I}}}{u' - u} \psi_{\hat{I}}(u),
\end{equation*}
where $\mathbf{X}_{J_0}^{\hat{I}}$ is the representation \eqref{eq:cla5}.

An order by order computation in \eqref{eq:boh} provides
\begin{equation}
\label{boh}
  \tilde{\psi}_{\hat{I}(p)}(u(p))\,=\,
e^{\frac{i}{2} \mathbf{X}_{J_0}^{\hat{I}}} \psi_{\hat{I}(p)}(u(p)),
\end{equation}
\ie{} the natural action of the chiral algebra on the vertex algebra
fields translates into the natural action of the representation of an
associated element of $G_r$ on the components of a given
$\hat{\mathfrak{g}}_1$-module primary field.

Let us now consider two adjacent cylindrical ends, \cyl{p} and
\cyl{q} together with the ribbon graph edge $\rho^1(p,q)$ which they 
share. Furthermore let us also assume that the theory on the $(p)$-th
polytope is deformed by the action of the boundary term $S_{B(p)} =
\int_0^{L(p)} du(p) \mathbf{J}_1(u(p))$, while the theory on the
$(q)$-th polytope is deformed by $S_{B(q)} = \int_0^{L(q)}
du^\prime(q) \mathbf{J}_2(u^\prime(q))$. According to
\eqref{eq:bs_schom}, the associated boundary states are defined as $
\Vert g_1 \Rangle = g_1 \, \Vert B \Rangle_{free} $ and $\Vert g_2
\Rangle \,=\, g_2 \, \Vert B \Rangle_{free}$.  In such a framework,
according to computations in section \ref{sec:bio}, in order to
characterize Boundary Insertion Operators it is sufficient to specify
the $(p,q)$ glueing automorphism entering in \eqref{eq:pq_cur}.

To this avail, the parametrization \eqref{eq:bs_schom} above introduced 
is not so efficient since it does not allow 
to successfully explain how the transition between pairwise adjacent
boundary conditions takes place.  

Thus we need to provide on $\partial$\cyl{p} a new representation for 
the infinite set of boundary conditions merging the choice of an element
within this set with the requirement to have a BIO acting
``\emph{\`a la Cardy}''.

As a first step in this direction we prove that Cardy boundary states are
those associated to deformations of the
unperturbed theory induced by elements in the center $B(G_r)$ of the
universal covering group $G_r$ generated by exponentiation of the 
parent finite algebra $\mathfrak{g}$.

It can be checked case by case that the center 
$B(G_r)$ is isomorphic to the group of outer automorphisms of the affine
algebra $\hat{\mathfrak{g}}$, namely $\mathcal{O}(\hat{\mathfrak{g}})$.
It is defined as a regular subgroup of the permutation group $D(\hat{
\mathfrak{g}})$ which is nothing but the symmetry group of 
$\hat{\mathfrak{g}}$ Dynkin diagram \cite{DiFrancesco}. Being 
$\hat{\mathfrak{g}}$ a simply laced affine algebra, the whole set can be 
explicitly classified as summarized in table 
\ref{tab:ADE1weights}. Let
us notice that the order of $\mathcal{O}(\hat{\mathfrak{g}})$, and
consequently of $B(G_r)$ coincides with the number of moduli of
$\hat{\mathfrak{g}}_k$ with $k=1$. Furthermore the isomorphism between
$\mathcal{O}(\hat{\mathfrak{g}})$ and $B(G_r)$ is realized 
associating to every element $A \in \mathcal{O}(\hat{\mathfrak{g}})$
the element $b \in B(G_r)\hookrightarrow G_r$:
\begin{equation}
  \label{eq:bA}
  b_A \,=\, e^{- 2 \pi i A \hat{\omega}_0\cdot H},
\end{equation}
where $A\omega_0 \in P_+^k(\hat{\mathfrak{g}})$ and $\hat{\lambda}
\cdot H =\sum\limits_i\lambda_i H_0^i - \lambda_0 L_0$.

\noindent Let us now choose \mbox{$\Vert B \Rangle_{free} \,\equiv\, \Vert 
\hat{\omega}_0\Rangle$} as a boundary state associated to the free theory.

The perturbation induced by a boundary
term associated to $b\in B(G_r)$ acts trivially on the glueing
condition because $b$, defined as in equation \eqref{eq:bA}, commutes
with all the affine algebra generators $J^a_n$. Thus
$\gamma_b(\overline{J}^a_{-n}) \,=\, \overline{J}^a_{-n}$ and the
correspondent rotated boundary state must satisfy the unperturbed
glueing condition, \ie{} it is a Cardy's boundary state.

Moreover, for each element $\hat{\omega}_i \in P_+^1$, it exists a unique
element $A_i \in \mathcal{O}(\hat{\mathfrak{g}})$ such that
\begin{equation}
  \label{eq:oi}
  \hat{\omega}_i \,=\, A_i
  \hat{\omega}_0.  
\end{equation}
Thus, to complete the proof of our statement, it is sufficient to show
that, it holds:
\begin{equation}
  \label{eq:cbsi}
  \Vert \hat{\omega}_i \Rangle \,=\, b_{A_i} 
  \Vert \hat{\omega}_0 \Rangle \quad 
  \text{with} \quad
  b_{A_i} \,=\, e^{- 2 \pi i A_i \hat{\omega}_0\cdot H} \,\in\,B(G_r).
\end{equation}
Hence let us introduce the following notation for Cardy's boundary 
states:
\begin{equation}
  \label{eq:cbsn}
  \Vert \hat{\omega}_I \Rangle
\,=\, \sum\limits_J\hat{\omega}_{I J} \vert \hat{\omega}_J \Rangle,
\end{equation}
where $\hat{\omega}_{IJ}\doteq
\frac{\mathcal{S}_{\hat{I}\hat{J}}^{ext}}
{\sqrt{\mathcal{S}_{\hat{0}\hat{J}}^{ext}}}$ and where 
$\vert \hat{\omega}_J\Rangle$ is the Ishibashi state built upon the  
$\hat{\omega}_J$-th module. 

All descendants in the module
$\mathcal{H}^{\hat{\mathfrak{g}}}_{\hat{\omega}_K}$ have the same
eigenvalue with respect to $b_{A_i}$ because the generators of the
algebra are unaffected by the action of the center:
\begin{equation*}
  b_{A_i} \vert \omega'\rangle \,=\, 
  e^{-2 \pi i (A_i\hat{\omega}_0,\hat{\omega}_K')} \vert \omega'\rangle \,=\,
  e^{-2 \pi i (A_i\hat{\omega}_0,\hat{\omega}_K)} \vert \omega'\rangle. 
  \quad \forall\,\vert \omega'\rangle \,\in\,
  \mathcal{H}^{\hat{\mathfrak{g}}}_{\hat{\omega}_I}
\end{equation*}
The same holds also for Ishibashi states, which are linear combinations 
of descendant states:
\begin{equation}
  \label{eq:bis}
  b_{A_i} \vert \hat{\omega}_J \Rangle \,=\, 
  e^{-2 \pi i (\hat{\omega}_i,\hat{\omega}_J)}
  \vert \hat{\omega}_J \Rangle 
  \,=\,
  \begin{cases}
    \vert \hat{\omega}_0 \Rangle & \text{if}\;J\,=\,0 \\
    e^{- 2 \pi i F_{i j}} 
    \vert \hat{\omega}_j \Rangle & \text{if}\;J\,=\,j \\
  \end{cases},
\end{equation}
where $F_{ij} = (\omega_i,\omega_j)$ is the quadratic form matrix of
the parent finite algebra. Since $(\omega_i,\omega_j) =
(\hat{\omega}_i,\hat{\omega}_j)$ the proof of \eqref{eq:cbsi} reduces to 
verify the following identity:
\begin{equation}
  \label{eq:prt}
  e^{- 2 \pi i (A_i \hat{\omega}_0 , \hat{\omega}_J)}\,
  \mathcal{S}_{\hat{0} \hat{J}}^{ext} \,=\, 
  \mathcal{S}_{\hat{i} \hat{J}}^{ext}.
\end{equation}

This last identity holds since the left hand side is the natural action 
of the automorphism $A_i\in\mathcal{O}(\hat{\mathfrak{g}})$ on the 
extended modular matrix:
\begin{equation*}
  e^{- 2 \pi i (A_i \hat{\omega}_0 , \hat{\omega}_J)}\,
  \mathcal{S}_{\hat{0} \hat{J}}^{ext} \,=\, 
  \mathcal{S}_{A_i(\hat{0}) \hat{J}}^{ext}\,=\,
  \mathcal{S}_{\hat{i} \hat{J}}^{ext}.
\end{equation*}

In view of this result, we can exploit the construction in
\eqref{eq:bs_schom} to parametrize the generic boundary condition
defined over the (inner or outer) boundary of the $k$-th cylindrical
end, represented by the boundary state $\Vert g(k) \Rangle$, with a
pair of elements:
\begin{equation}
  \label{eq:boundary_par}
  \left(
\Vert \hat{\omega}_I \Rangle,\, \Gamma(k)
\right), \qquad \text{with}\quad
\begin{cases}
\hat{\omega}_I & \,\in\, P_+^1(\hat{\mathfrak{g}}) \\
\Gamma(k) & \,\in\,\frac{G_r}{B(G_r)}
\end{cases}
\end{equation}
being $\Vert \hat{\omega}_I \Rangle$ a Cardy's boundary state and 
$\Gamma(k)\in\frac{G_r}{B(G_r)}$ such that:
\begin{equation}
\label{eq:poly_bs}
  \Vert g(k) \Rangle
  \,=\,
  \Gamma(k) \, \Vert \hat{\omega}_J(k) \Rangle.
\end{equation}

We can show that this parametrization is not only a formal datum. As a
matter of fact coset theory ensures that, $\forall\,g\,\in\,G_r$, we
can choose a representative $\Gamma \in \frac{G_r}{B(G_r)}$ and
an element $b_I \in B(G_r)$ such that $g$ is uniquely decomposed
as
\begin{equation}
\label{eq:deco}
  g\,=\,\Gamma\,\cdot\,b_I.
\end{equation}

Moreover, being $\frac{G_r}{B(G_r)}$ a Lie group\footnote{ The
  converse to Lie third theorem \cite{Gilmore} ensures that, given a
  finite dimensional abstract real Lie algebra $\mathfrak{g}$, there
  is a single \emph{simply connected} Lie group G whose Lie algebra is
  isomorphic to $\mathfrak{g}$, namely the \emph{universal covering
    group} generated by $\mathfrak{g}$. All other groups with the same
  Lie algebra can be obtained from the universal covering one by
  quotient with one of its invariant discrete subgroups - say $D$.

  The factor group $H=\frac{G}{D}$ is a multiply connected Lie group
  since, quoting from \cite{Fulton-Harris}, it holds:
  \begin{proposition}
    \label{pr:Lie_str}
    Let $G$ be a Lie group with center $B(G)$ such that $D \subseteq B(G)$ 
    is a finite subgroup of $G$. Then there is a unique Lie group structure
    on the
    quotient group $H = G/D$ such that the quotient map $G \rightarrow
    H$ is a Lie group map.
  \end{proposition}},
uniqueness of \eqref{eq:deco} allows us to define a 
global smooth map:
\begin{center}
  \begin{tabular}{cccccc}
    $\sigma$: \quad & $\frac{G_r}{B(G_r)}$ &
    $\longrightarrow$ &
    $\frac{G_r}{B(G_r)} \times B(G_r)$ & $\hookrightarrow$ & $G_r$ \\
    & $\Gamma$ & $\mapsto$ & $(\Gamma,\,e)$ & $\mapsto$ &
    $\Gamma\,\cdot\,e$
  \end{tabular},
\end{center}
where $e$ is the identity of $G_r$. Hence, since all the hypotheses of
proposition 1 are met and since $\sigma$ is also a group homomorphism,
we can consider
$\left(\frac{G_r}{B(G_r)},\,\sigma\right)$ a Lie subgroup of $G_r$. The
inclusion $\sigma:\frac{G_r}{B(G_r)}\hookrightarrow G_r$ translates 
at a level of Lie algebras as
\begin{equation}
  \label{eq:isoal}
  d \sigma\,: \quad \mathfrak{g} \;\rightarrow\, \mathfrak{g}.
\end{equation}

\noindent The following holds \cite{Warner}:
\begin{proposition}
  \label{p:em}
  Let $(H,\sigma)$ be a Lie subgroup of $G_r$ with Lie algebra
  $\mathfrak{h}$ and let
  $X\in\mathfrak{g}$. If $X \in d \sigma (\mathfrak{h})$, then $e^{t
    X} \in \sigma(H)$ for all $t\in\mathbb{R}$. Conversely, if $e^{t X} \in
  \sigma(H)$ for $t$ in some open real interval, then $X \in d \sigma
  (\mathfrak{h})$.
\end{proposition}
(Proof can be found at pg. 104 chapter 3 in \cite{Warner}).

According to this last proposition, the existence of a (global)
exponential map from the image of \eqref{eq:isoal} into 
$\frac{G_r}{B(G_r)}$ is granted, {\it i.e.} for any $\Gamma \in 
\frac{G_r}{B(G_r)}$, we can uniquely write 
the immersion $\sigma(\Gamma)\in G_r$ as
\begin{equation*}
 \sigma(\Gamma)\equiv\Gamma \,=\, e^{i\sum_a \Gamma_a J^a},
\end{equation*}
where $\{J^a\} \doteq \{E^{\alpha}_0,\,H^r_0\}$ are the
(Cartan-Weyl) generators
of the horizontal subalgebra of $\hat{\mathfrak{g}}_1$ 
isomorphic to $\mathfrak{g}$. 
Let us denote the element $b_I \in B(G_r)$ as $b_I
\doteq e^{ib_r H^r_0}$; the Baker-Campbell-Hausdorff
formula \cite{Gilmore, Fulton-Harris} ensures that it holds a precise
relation among coefficients $b_r$, $\Gamma_a$ and $g_a$ such that we
can write:
\begin{equation}
  \label{eq:bhfc}
  g \,=\, e^{i\sum_a g_a J^a_0} \,=\, 
  e^{i\sum_a \Gamma_a J^a}\,e^{ib_r H^r_0}.
\end{equation}
The associated boundary state $\Vert g \Rangle$ can be uniquely
written as
\begin{equation*}
  \Vert g(k) \Rangle \,=\, g\,\Vert \hat{\omega}_0 \Rangle 
  \,=\, \Gamma \cdot b_I \, \Vert \hat{\omega}_0 \Rangle 
  \,=\, \Gamma \,\Vert \hat{\omega}_I \Rangle.
\end{equation*}

Hence we have split the deformation process in
\eqref{eq:pa}-\eqref{eq:bs_schom} in two subsequent steps. The first
involves a deformation induced by a boundary action term $\mathcal{S}_b=
\int du(p)\,\left[ \sum_r b_r H^r(u(p)) \right]$, uniquely determined by
a group element in the center of the universal covering group,
$B(G_r)$. This deformation actually maps the old free boundary state
into a Cardy one, while its action changes the boundary operators
only for a multiplication of their components by a constant phase
factor.

The second step is instead a deformation induced by the boundary term
$\mathcal{S}_{\Gamma}(p) = \int du(p)\,\sum_a \Gamma_a
J^a(u(p))$, which acts on a Cardy boundary state
mapping it into $\Vert g \Rangle \,=\, \Gamma \,\Vert
\hat{\omega}_I \Rangle$; at the same time it can act non-trivially on 
boundary operators.

To summarize, the above parametrization states that we are actually
performing a deformation of $\hat{\mathfrak{g}}_1$-WZW model described
``\`a la Cardy'' by means of a boundary term such that the associated
group element $e^{i\sum_a \Gamma_a J^a_0}$ is the image of
$\Gamma\,\in\,\frac{G_r}{B(G_r)}$ in $G_r$ by means of the map
$\sigma$.

Within this framework, the amplitude intermediate channels associated
with the generic $(p,q)$-edge of the ribbon graph, will correspond
to an automorphism induced by the operator $g_1(p)g_2^{-1}(q)$.
Parametrization \eqref{eq:boundary_par} allows to ``make explicit'' the
action of this glueing automorphism \eqref{eq:pq_cur} with two
separate objects: a map which relates dynamically the two Cardy
boundary states and an action on the residues $\frac{G_r}{B(G_r)}$
deformations induced by $\Gamma_1$ and $\Gamma_2$. The former
is easily retrieved reasoning in analogy with the definition of boundary
conditions changing operators of rational minimal models. As a matter
of fact we can define this ``first act'' of the glueing process as the
fusion between the representations associated to the two adjacent
Cardy's boundary states and the one a BIO carries. Thus, let us consider
the $(p,q)$ edge $\rho^1(p,q)$, and boundary
conditions specified uniquely by the central actions $S_{b}(p)$ and
$S_{b}(q)$. If $\Vert \hat{\omega}_J(p) \Rangle$ and $\Vert
\hat{\omega}_J(q) \Rangle$ are the associated boundary states, which
are shared by $\rho^1(p,q)$, BIOs on $\rho^1(p,q)$ are defined as
\begin{equation}
  \label{eq:su2bio}
  \psi_{\hat{I}(p,q)}^{\hat{J}(p)\,\hat{J}(q)} (x(p,q)) \,=\,
\mathcal{N}_{\hat{J}(p)\,\hat{I}(p,q)}^{\hat{J}(q)}\,
\psi_{\hat{I}(p,q)}(x(p,q)),
\end{equation}
\ie{} they are the $\hat{g}_{k=1}$ primary fields weighted by the
fusion rule ${N}_{\hat{J}(p)\,\hat{I}(p,q)}^{\hat{J}(q)}$. These
are provided by a combination of the $\mathcal{S}^{ext}$ matrix entries
via the Verlinde formula
\begin{equation}
  \label{eq:verlfor}
  \mathcal{N}_{\hat{J}(p)\,\hat{I}(p,q)}^{\hat{J}(q)} \,=\,
  \sum_{\hat{\omega}_K \in P_+^1(\hat{\mathfrak{g}})}
  \frac{S^{ext}_{\hat{J}(p)\,\hat{K}}\,S^{ext}_{\hat{I}(p,q)\,\hat{K}}\,
\overline{S}^{ext}_{\hat{K}\,\hat{J}(q)}}{S^{ext}_{\hat{0}\,\hat{K}}}.
\end{equation}

At this stage the ``second act'' is straightforward. Let us switch
on the boundary terms $S_{\Gamma_1}(p)$ and $S_{\Gamma_2}(q)$
on the inner boundaries of \cyl{p} and \cyl{q}. We can deform the
$(p,q)$ theory with a suitable combination of currents which maps 
the $S_{\Gamma_1}(p)$-induced deformation into the
$S_{\Gamma_2}(q)$-induced deformation. In the forthcoming analysis, we will
show that this choice allows the two ends to glue dynamically in such a
way that such a dynamic is actually governed by the fusion rules of the
WZW model.
The explicit expression of this combination of currents is
established by requiring that the image into $G_r$ of the associated
group element in
$\frac{G_r}{B(G_r)}$ is $\overline{\Gamma} \,=\,
\Gamma_2\Gamma_1^{-1}$. Viceversa, if we consider the $(q,p)$ theory - 
formally distinct form the $(p,q)$-one -, the image of
the associated element would be $\overline{\Gamma}^{-1} \,=\,
\Gamma_1\Gamma_2^{-1}$. Thus let us write the desired defect term as
\begin{equation}\label{eq:cla7}
  S_{(p,q)} \,=\,
  \int\limits_{\rho^1(p,q)} d x(p,q) \, \sum_a^{\dim\mathfrak{g}} 
  \overline{\Gamma}_a \mathbb{J}_{a\,(p,q)}(x(p,q)),  
\end{equation}
where $\mathbb{J}_{a\,(p,q)}$ is defined as in \eqref{eq:pq_cur}.  
The combination of \eqref{eq:pq_cur} and \eqref{eq:1al} allows to write
the above defect term exactly as a boundary perturbing term for the
\cyl{p} theory (let us remember that the formal expression in
\eqref{eq:pq_cur} has to be defined separately for the holomorphic and
antiholomorphic components): it maps the boundary state in $\Gamma_1
\,\Vert \hat{\omega}_I\Rangle$ into $\Gamma_2 \,\Vert
\hat{\omega}_I\Rangle$.

To describe the effect of \eqref{eq:cla7} on Boundary Insertion
Operators, let us consider the functional expression of their
components, dropping the dependence from the fusion rule factor:
\begin{equation*}
  \psi_{[\hat{I},m](p,q)},\qquad  \text{with}\quad
\hat{\omega}_I\,\in\,P_+^1.\quad
m\,=\,1,\,\ldots,\,\text{dim}|\hat{\omega}_I| 
\end{equation*}

Since functional and conformal properties of Boundary Insertion
Operators are strictly analogue to those of ordinary boundary operators
(see section \ref{sec:bio}), we can apply to the formers exactly the same
arguments as in equation \eqref{eq:boh} and subsequents. Hence the
defect term will deform BIOs by means of a
rotation:
\begin{equation*}
  \psi_{[\hat{I},m](p,q)} \,\longrightarrow\,
  e^{\frac{i}{2} X_{\overline{\Gamma}}^{\hat{I}}} \,
  \psi_{[\hat{I},m](p,q)},
\end{equation*}
\ie{} the action of the chiral algebra translates into the action of
the associated group via its unitary representations. Thus, restoring 
the fusion coefficients, we have
the following expression for boundary insertion operators in the
rational limit of the conformal theory:
\begin{equation}
  \label{eq:final_bio}
  \psi^{[\hat{J}_2,\,\Gamma_2](q)\,[\hat{J}_1,\,\Gamma_1](p)}
  _{[\hat{I},\,m](p,q)}
  \,=\,
  \sum_{n=0}^{\text{dim}|\hat{I}|}  
R^{\hat{I}(p,q)}_{m\,n(p,q)}(\Gamma_2{\Gamma_1}^{-1}) \,
  \psi_{[\hat{I},\,n](p,q)}^{\hat{J}_2(q)\,\hat{J}_1(p)},
\end{equation}
where $\psi_{\hat{J}(p,q)}^{\hat{J}(p)\,\hat{J}(q)} (x(p,q)) \,=\,
\mathcal{N}_{\hat{J}(p)\,\hat{I}(p,q)}^{\hat{J}(q)}
\,\psi_{\hat{I}(p,q)}(x(p,q))$ and where $R^{\hat{I}(p,q)}_{m\,n(p,q)}=
\exp\left[\frac{i}{2}X^{\hat{I}(p,q)}\right]_{m\,n(p,q)}$ being
$X^{\hat{I}}$ the operator introduced in \eqref{eq:cla5}.

\subsection{The algebra of Boundary Insertion Operators}
\label{sec:algebrabio}

The aim of this rather technical section is to show that, with
boundary insertion operators defined as in equation
\eqref{eq:final_bio}, boundary perturbations do not affect
the algebra of boundary operators which is completely fixed in terms
of the fusion rules of the WZW-model. This is indeed a check that our
prescription for the $(p,q)$ glueing automorphism and its consequent
action on BIOs is consistent: as a matter of
fact all deformations we have introduced are actually truly marginal
ones and, thus, they must not break the chiral symmetry generated by
$\hat{\mathfrak{g}}$.

The algebra of rotated BIOs follows from their definition. Let us
notice that rotated BIOs are just a superposition of the different
components of Cardy $\hat{\mathfrak{g}}_1$-chiral primary operators'
components.

Let us focus our attention on the two-point function  between a
$p$-to-$q$ and $q$-to-$p$ mediating operators\footnote{Let us remember
that the other possible two-points function loses its physical meaning
after a suitable assignation of Chan-Paton factors (see comments at
the end of section \ref{sec:osgauge}).}.
 We have to compute:
 \begin{multline}
   \left\langle
    \psi^{[\hat{J}_2,\,\Gamma_2](q)\,[\hat{J}_1,\,\Gamma_1](p)}
    _{[\hat{I},\,m](p,q)}
    (x_1(p,q)) \,
    \psi^{[\hat{J}_3,\,\Gamma_3](p)\,[\hat{J}_4,\,\Gamma_4](q)}
    _{[\hat{I}',\,m'](q,p)}
    (x_2(q,p))
  \right\rangle.
 \end{multline}

As a first step we must notice that a coherent glueing imposes the two
operators to mediate between the same boundary conditions (see
equation \eqref{eq:2points}). Accordingly the above expression reduces to:
\begin{multline}
  \label{eq:rot2points2}
  \left\langle
    \psi^{[\hat{J}_2,\,\Gamma_2](q)\,[\hat{J}_1,\,\Gamma_1](p)}
    _{[\hat{I},\,m](p,q)}
    (x_1(p,q)) \,
    \psi^{[\hat{J}_1,\,\Gamma_1](p)\,[\hat{J}_2,\,\Gamma_2](q)}
    _{[\hat{I}',\,m'](q,p)}
    (x_2(q,p))
  \right\rangle\,=\,\\
  \sum_{n\,n'}
  R^{\hat{I}(p,q)}_{m\,n(p,q)}(\Gamma_2{\Gamma_1}^{-1})
  R^{\hat{I}'(q,p)}_{m'\,n'(q,p)}(\Gamma_1{\Gamma_2}^{-1}) \times \\
  \left\langle
    \psi^{\hat{J}_2(q)\,\hat{J}_1(p)}_{[\hat{I},\,n](p,q)}(x_1(p,q)) \,
    \psi^{\hat{J}_1(p)\,\hat{J}_2(q)}_{[\hat{I}',\,n'](q,p)}(x_2(q,p))
  \right\rangle.
\end{multline}

Let us notice that, in the previous expression, we are dealing with a
representation of the diagonal subgroup of the direct product
$\frac{G_r}{B(G_r)}(p,q)\times\frac{G_r}{B(G_r)}(q,p)$; hence it holds
(see eq. \eqref{eq:dir_prod}):
\begin{equation}
  \label{eq:dirprod_rep}
  R^{\hat{I}(p,q)}_{m\,n(p,q)}(\Gamma_2{\Gamma_1}^{-1})
  R^{\hat{I}'(q,p)}_{m'\,n'(q,p)}(\Gamma_1{\Gamma_2}^{-1})
  \,=\,R^{\hat{I} \times \hat{I}'}_{m\,n;\,m'\,n'}(\mathbb{I}).
\end{equation}

The Clebsh-Gordan expansion (eq. \eqref{eq:CG_series}) gives
(we omit the polytope indexes in the Clebsh-Gordan coefficients):
\begin{multline}
  \left\langle
    \psi^{[\hat{J}_2,\,\Gamma_2](q)\,[\hat{J}_1,\,\Gamma_1](p)}
    _{[\hat{I},\,m](p,q)}
    (x_1(p,q)) \,
    \psi^{[\hat{J}_1,\,\Gamma_1](p)\,[\hat{J}_2,\,\Gamma_2](q)}
    _{[\hat{I}',\,m'](q,p)}
    (x_2(q,p))
  \right\rangle\,=\,\\
  \sum_{n\,n'}\,\sum_{\hat{J}\,N} 
  C_{\hat{I}\,m\,\hat{I}^\prime\,m^\prime}^{\hat{J}\,N}
  C_{\hat{I}\,n\,I^\prime\,n^\prime}^{\hat{J}\,N}
  \left\langle
    \psi^{\hat{J}_2(q)\,\hat{J}_1(p)}_{[\hat{I},\,n](p,q)}(x_1(p,q)) \,
    \psi^{\hat{J}_1(p)\,\hat{J}_2(q)}_{[\hat{I}',\,n'](q,p)}(x_2(q,p))
  \right\rangle\,=\,\\
  \left\langle
    \psi^{\hat{J}_2(q)\,\hat{J}_1(p)}_{[\hat{I},\,m](p,q)}(x_1(p,q)) \,
    \psi^{\hat{J}_1(p)\,\hat{J}_2(q)}_{[\hat{I}',\,m'](q,p)}(x_2(q,p))
  \right\rangle,
\end{multline}
where, in the last equation, we have used the completeness of Clebsh-Gordan
coefficients (see equation \eqref{eq:CG_unit2}).

\begin{figure}[!t]
  \centering
  \includegraphics[width=.5\textwidth]{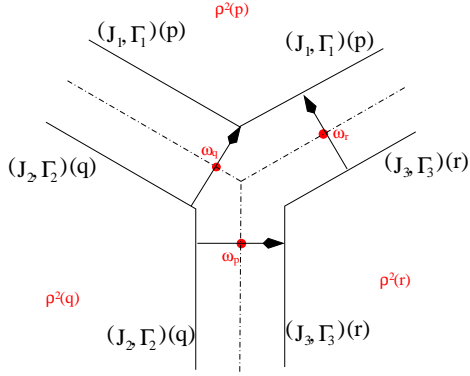}
  \caption{OPE between rotated Boundary Insertion Operators.}
  \label{fig:BIO_ope-deformed}
\end{figure}

To calculate the OPE of rotated BIOs, let us notice that the rotation
generated by the boundary condensate does not change the coordinate
dependence. Let us consider the situation depicted in figure
\ref{fig:BIO_ope-deformed}.

OPE between
$\psi^{[\hat{J}_1,\,\Gamma_1](p)\,[\hat{J}_3,\,\Gamma_3](r)}
_{[\hat{I}_1,\,m_1](r,p)}$ and
$\psi^{[\hat{J}_3,\,\Gamma_3](r)\,[\hat{J}_2,\,\Gamma_2](q)}
_{[\hat{I}',\,m'](q,r)}$ will mediate a change in boundary conditions
from $[\hat{J}_2,\,\Gamma_2](q)$ to $[\hat{J}_1,\,\Gamma_1](p)$. In
particular,
\begin{multline}
  \notag
  \psi^{[\hat{J}_1,\,\Gamma_1](p)\,[\hat{J}_3,\,\Gamma_3](r)}
  _{[\hat{I}_1,\,m_1](r,p)}
  (\omega_r) \,
  \psi^{[\hat{J}_3,\,\Gamma_3](r)\,[\hat{J}_2,\,\Gamma_2](q)}
  _{[\hat{I}_2,\,m_2](q,r)}
  (\omega_q)\,=\, \\
  \sum_{n_1(r,p)\,n_2(q,r)}\,
  R^{\hat{I}_1(r,p)}_{m_1\,n_1(p,q)}(\Gamma_1{\Gamma_3}^{-1})\,
  R^{\hat{I}_2(q,r)}_{m_2\,n_2(q,r)}(\Gamma_3{\Gamma_2}^{-1})\,
  \psi^{\hat{J}_1(p)\,\hat{J}_3(r)}_{[\hat{I}_1,\,n_1](r,p)}(\omega_r) \,
  \psi^{\hat{J}_3(r)\,\hat{J}_2(q)}_{[\hat{I}_2,\,m_2](q,r)}(\omega_q).
\end{multline}

We are dealing again with a representation of the diagonal subgroup
of the direct product
$\frac{G_r}{B(G_r)}(r,p)\times\frac{G_r}{B(G_r)}(q,r)$;
hence, applying \eqref{eq:dir_prod} and the Clebsh-Gordan series
expansion \eqref{eq:CG_series}, we are left with
\begin{multline}
  \label{interm1}
  \sum_{\substack{n_1(r,p)\\n_2(q,r)}}\,
  \sum_{\hat{I}}\sum\limits_{m,n = 1}^{\dim\hat{I}}
  C^{\hat{I}\,m}_{\hat{I}_1(r,p)\,m_1(r,p)\,\hat{I}_2(q,r)\,m_2(q,r)}\,
  R^{\hat{I}}_{m\,n}(\Gamma_1\Gamma_2^{-1})\, \\ \times
  C^{\hat{I}\,n}_{\hat{I}_1(r,p)\,n_1(r,p)\,\hat{I}_2(q,r)\,n_2(q,r)}\,
  \psi^{\hat{J}_1(p)\,\hat{J}_3(r)}_{[\hat{I}_1,\,n_1](r,p)}(\omega_r) \,
  \psi^{\hat{J}_3(r)\,\hat{J}_2(q)}_{[\hat{I}_2,\,n_2](q,r)}(\omega_q).
\end{multline}

According to \eqref{eq:ope1}, the OPE between undeformed Boundary
Insertion Operators reads:
\begin{multline}
  \psi^{\hat{J}_1(p)\,\hat{J}_3(r)}_{[\hat{I}_1,\,n_1](r,p)}(\omega_r) \,
  \psi^{\hat{J}_3(r)\,\hat{J}_2(q)}_{[\hat{I}_2,\,n_2](q,r)}(\omega_q)\,=\,
  \sum_{\hat{I}_3, n_3}\left\vert
    \omega_r\,-\,\omega_q
  \right\vert^{H(q,p)\,-\,H(r,p)\,-\,H(q,r)}\\
  C^{\hat{I}_3\,n_3}_{\hat{I}_1\,n_1\,\hat{I}_2\,n_2}\,
  \mathcal{C}_{\hat{I}_1\,\hat{I}_2\,\hat{I}_3}
  ^{\hat{J}_1(p)\,\hat{J}_3(r)\,\hat{J}_2(q)}\,
  \psi^{\hat{J}_1(p)\,\hat{J}_2(q)}_{[\hat{I}_3,\,n_3](q,p)}(\omega_q),
\end{multline}
where the Clebsh-Gordan coefficients
$C^{\hat{I}_3\,n_3}_{\hat{I}_1\,n_1\,\hat{I}_2\,n_2}$ compensate the
fact that the l.h.s. and r.h.s. terms have different transformation behavior
under the action of the horizontal $\mathfrak{g}$ algebra, while the
coefficients $\mathcal{C}_{\hat{I}_1\,\hat{I}_2\,\hat{I}_3}
^{\hat{J}_1(p)\,\hat{J}_3(r)\,\hat{J}_2(q)}$ reflect the non trivial
dynamic on each trivalent vertex of the ribbon graph.

The inclusion of this last OPE into \eqref{interm1} and the Clebsh-Gordan
coefficients unitarity (equation \eqref{eq:CG_unit2}) leaves us with:
\begin{multline}
  \label{eq:rotated_ope}
  \psi^{[\hat{J}_1,\,\Gamma_1](p)\,[\hat{J}_3,\,\Gamma_3](r)}
  _{[\hat{I}_1,\,m_1](r,p)}
  (\omega_r) \,
  \psi^{[\hat{J}_3,\,\Gamma_3](r)\,[\hat{J}_2,\,\Gamma_2](q)}
  _{[\hat{I}_2,\,m_2](q,r)}
  (\omega_q)\,=\, \\
  \sum_{j_3\,m}
  C^{\hat{I}_3\,m}_{\hat{I}_1\,m_1\,\hat{I}_2\,m_2}\,
  \mathcal{C}_{\hat{I}_1\,\hat{I}_2\,\hat{I}_3}
  ^{\hat{J}_1(p)\,\hat{J}_3(r)\,\hat{J}_2(q)}\,
  \psi^{[\hat{J}_1,\,\Gamma_1](p)\,[\hat{J}_2,\,\Gamma_2](q)}
  _{[\hat{I}_3,\,m_3](q,p)}
  (\omega_p).
\end{multline}
Thus, we demonstrated that OPE between rotated BIOs is formally equal
to OPE between unrotated ones. Accordingly, on the ribbon graph the non
trivial dynamic is given by the fusion among the three representations
entering in each trivalent vertex.

This allows us to further pursue our investigation and to consider the 
four-points function between BIOs included on graph edges which are 
among four adjacent polytopes:
\begin{equation}
  \label{eq:4pf}
  \langle
  \psi_{\hat{I}_1(s,p)}^{\hat{J}_1(p)\,\hat{J}_4(s)}\,
  \psi_{\hat{I}_2(r,s)}^{\hat{J}_4(s)\,\hat{J}_3(r)}\,
  \psi_{\hat{I}_3(q,r)}^{\hat{J}_3(r)\,\hat{J}_2(q)}\,
  \psi_{\hat{I}_4(p,q)}^{\hat{J}_2(q)\,\hat{J}_1(p)}
  \rangle.
\end{equation}

The variable connectivity of the triangulation becomes fundamental in
this computation since it allows to state a correspondence between
the two possible factorizations out of which we can compute
\eqref{eq:4pf} and the two ways we can fix adjacency of the four
polytopes involved in the analysis.

\begin{figure}[!t]
  \centering
  \includegraphics[width=\textwidth]{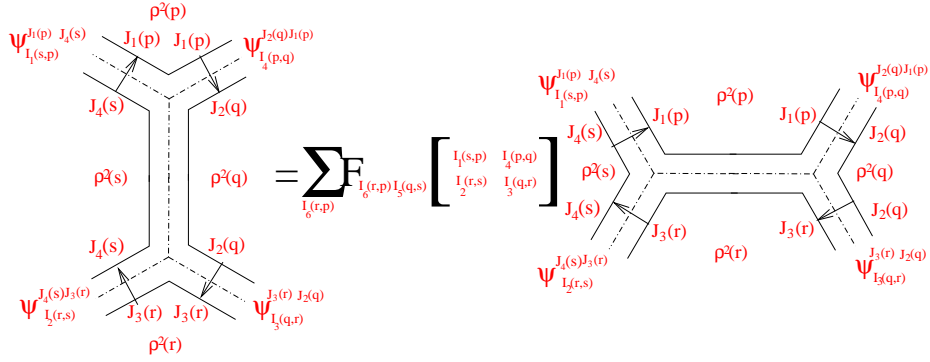}
  \caption{Four-points function crossing symmetry.}
  \label{fig:BIO_4points}
\end{figure}

Let us consider the natural picture in which we construct a 
four-points function arises, namely the neighborhood of two near trivalent
vertexes. Due to the variable connectivity of the triangulation, the
two configurations shown in figure \ref{fig:BIO_4points} are both
admissible. The transition from the situation depicted in the l.h.s. and
the one in the r.h.s. of the pictorial identity in figure
\ref{fig:BIO_4points} corresponds exactly to the transition between
the $s$-channel and the $t$-channel of the four-points blocks of a
single copy of the bulk theory.

The two factorizations of the above four-points function are related by 
the bulk crossing matrices:
\begin{equation}
\label{fusion_matrices}
F_{\hat{I}_6(s,q)\,\hat{I}_5(r,p)}
\begin{bmatrix}
  \hat{I}_4(p,s) & \hat{I}_1(q,p) \\
  \hat{I}_3(s,r) & \hat{I}_2(r,q)
\end{bmatrix}.
\end{equation}

\noindent The explicit computation of the two factorizations leads to the
relation
\begin{multline}
  \label{eq:fact}
  \mathcal{C}^{\hat{J}_4(s)\,\hat{J}_3(r)\,\hat{J}_2(q)}
  _{\hat{I}_2(r,s)\,\hat{I}_3(q,r)\,\hat{I}_5(q,s)}\,
  \mathcal{C}^{\hat{J}_1(p)\,\hat{J}_4(s)\,\hat{J}_2(q)}
  _{\hat{I}_1(s,p)\,\hat{I}_5(q,s)\,\hat{I}_1(s,p)}\,
  \mathcal{C}^{\hat{J}_1(p)\,\hat{J}_2(q)\,\hat{J}_1(p)}
  _{\hat{I}_1(s,p)\,\hat{I}_1(s,p)\,0}
  \,=\,\\
  \sum_{\hat{I}_5(r,p)}
  F_{\hat{I}_6(s,q)\,\hat{I}_5(r,p)}
  \begin{bmatrix}
    \hat{I}_4(p,s) & \hat{I}_1(q,p) \\
    \hat{I}_3(s,r) & \hat{I}_2(r,q)
  \end{bmatrix}\,\times\\
  \mathcal{C}^{\hat{J}_1(p)\,\hat{J}_4(s)\,\hat{J}_3(r)}
  _{\hat{I}_1(s,p)\,\hat{I}_2(r,s)\,\hat{I}_6(r,p)}\,
  \mathcal{C}^{\hat{J}_3(r)\,\hat{J}_2(q)\,\hat{J}_1(p)}
  _{\hat{I}_3(q,r)\,\hat{I}_4(p,q)\,\hat{I}_6(p,r)}\,
  \mathcal{C}^{\hat{J}_1(p)\,\hat{J}_3(r)\,\hat{J}_1(p)}
  _{\hat{I}_6(r,p)\,\hat{I}_6(p,r)\,0},
\end{multline}
\ie{} the usual BCFT sewing relation among boundary operators OPEs.

This statement completes our analysis of the conformal properties of the full
theory arising glueing together the BCFTs defined over each
cylindrical end; within the above construction BIOs play exactly the
same role as the usual boundary operators in BCFT.

This analogy allows us to apply to BIOs all boundary operators
properties. In particular, we can identify their OPE coefficients
describing interactions in the neighborhood of the $(p,q,s)$ vertex
of the ribbon graph with the fusion matrices \eqref{fusion_matrices}
with the following entries assignations:
\begin{equation}
  \label{eq:OPE_fusion}
  \mathcal{C}^{\hat{J}_1(p)\,\hat{J}_2(q)\,\hat{J}_3(s)}
  _{\hat{I}_1(q,p)\,\hat{I}_2(s,q)\,\hat{I}_3(s,p)}
  \,=\,
  F_{\hat{J}_2(q)\,\hat{I}_3(s,p)}
  \begin{bmatrix}
    \hat{J}_1(p)   &  \hat{J}_3(s)   \\
    \hat{I}_1(q,p) &  \hat{I}_2(s,q)
  \end{bmatrix}.
\end{equation}

Relation \eqref{eq:OPE_fusion}, first obtained  in \cite{Runkel:1998pm}
for the $A$-series minimal models, has been
recast for all minimal models and extended rational conformal field
theories in \cite{Behrend:1999bn} and \cite{Felder:1999ka} exploiting
the full analogy between equation \eqref{eq:fact} and the pentagonal
identity for the fusing matrices.

According to \cite{Alvarez-Gaume:1988vr}, WZW-models fusion matrices
coincide with the $6j$-symbols of the corresponding quantum group
with deformation parameter given by the $(k\,+\,g^\vee)$-th root of
the identity, where $k$ and $h^\vee$ are respectively the level and
the dual Coxeter number of the extended algebra (the list of dual
Coxeter numbers for the rank-$r$ simply laced algebras can be found in
table \ref{tab:ADE1weights}). Thus, with $k=1$, the OPEs coefficients
are the quantum group $G_{Q\,=\,e^{\frac{2 \pi i}{1 + h^\vee}}}$
$6j$-symbols:
\begin{equation}
  \label{eq:OPE_6j}
  \mathcal{C}^{\hat{J}_1(p)\,\hat{J}_2(q)\,\hat{J}_3(s)}
  _{\hat{I}_1(q,p)\,\hat{I}_2(s,q)\,\hat{I}_3(s,p)}
  \,=\,
  \begin{Bmatrix}
    \hat{I}_1(q,p) & \hat{J}_1(p) & \hat{J}_2(q)  \\
    \hat{J}_3(s) & \hat{I}_2(s,q) & \hat{I}_3(s,p)
  \end{Bmatrix}_{Q\,=\,e^{\frac{2 \pi i}{1 + h^\vee}}}.
\end{equation}

\section{Discussion and conclusions}

In this paper we have fully characterized the local coupling between a
scalar Rational Boundary Conformal Field Theory and a special class of
open surfaces $M_\partial$. The latter arise as uniformizations of a
Random Regge Triangulation. In this connection, the results in
previous section provide the main ingredient needed to write a
worldsheet amplitude defined over the full $M_\partial$. To this aim,
a possible candidate is a construction introduced in \cite{Carfora2}.
Exploiting an edge vertex factorization of the most general BIOs'
correlator we can write on the ribbon graph, we can express the
contribution to the graph-amplitude given by each set of $r$ fields
$\{X^i\}, i=1,\ldots,r$ associated to each factor entering in
\eqref{eq:sgrup} as
\begin{multline}
  Z(\Gamma, r) \,=\,\\
  \sum_{\{\hat{I}(r,p)\} \in P_1^+{\hat{\mathfrak{g}}}}
  \prod_{\rho^{0}(p,q,r)=1}^{N_{2}(T)} \mathcal{C}^{\hat{J}_1(p)
    \hat{J}_3(r) \hat{J}_2(q)}_{\hat{I}_1(r,p) \hat{I}_2(q,r)
    \hat{I}_3(q,p)}
  \times \\
  \times \prod_{\rho ^{1}(p,r)=1}^{N_{1}(T)}\left(
    b_{\hat{I}_1(r,p)}^{\hat{J}_1(p) \hat{J}_3(r})\right)
  ^{2}L(p,r)^{-2H_{\hat{I}_1(r,p)}}.
\end{multline}
The sum runs over all the $N_1(T)$ primaries of the chiral algebra
decorating the ribbon graph edges through the insertion of BIOs, and
with the OPE coefficients $\mathcal{C}^{\hat{J}_1(p) \hat{J}_3(r)
\hat{J}_2(q)}_{\hat{I}_1(r,p) \hat{I}_2(q,r) \hat{I}_3(q,p)}$ being
replaced by the associated $6j$ symbols.

Afterward each contribution must be applied on the associated $N_0(T)$
channels defined by the cylinder amplitude for the correspondent
directions.  Concerning any but fixed factor in eq. \eqref{eq:sgrup},
we want to define the transition amplitude between two boundary states
$\Vert g_1 \Rangle$ and $\Vert g_2 \Rangle$, the latter being
constructed out of the action of an element $g \in G_r$ on the first
one: $\Vert g_2 \Rangle \,=\, g\,\Vert g_1 \Rangle$. As proved in
details in section 4 of \cite{Recknagel:1998ih} or
\cite{Gaberdiel:2001xm}, it is easy to show that the amplitude
$\mathcal{A}_{\text{\cyl{p}}}^{g_1,\,g \cdot g_1}$ depends only upon
the conjugacy classes of $g \in G_r$ .  Therefore, we can choose to
deform the boundary state with an element in the maximal torus of
$G_r$, $h \,=\, e^{i\sum_{i=1}^r \lambda_iH^i}$.  Thus, if we
choose $\Vert g_1 \Rangle$ to coincide with one of the Cardy boundary states
$\Vert\hat{\omega}_K \Rangle$\footnote{Let us notice that, in view of
  results in section \ref{sec:algebrabio}, such a choice does not
  impose any restriction on the dynamic of the model.}, the amplitude
will involve a sum over $\hat{\mathfrak{g}}_1$ characters, twisted by
the action of $g\in G_r$:
\begin{equation*}
  \mathcal{A}_{\text{\cyl{p}}}^{\hat{K},\,g(\hat{K})} \,=\,
\sum_{\hat{I}\in P_+^1} \mathcal{N}_{\hat{K}\,\hat{K}}^{\hat{I}}
\text{Tr}_{\mathcal{H}_{\hat{I}}}[\tau_{h}q^{L_0^{(O)} - \frac{r}{24}}],
\end{equation*}
where $\tau_{h}$ is the action induced by the selected group element
on $\mathcal{H}_{\hat{I}}$.  Hence the full amplitude on a fixed
geometry parametrized by a choice of the ribbon graph $\Gamma$ and of
the set of localized curvature assignations $\{\varepsilon(s)\}, s =
1, \ldots N_0(T)$, becomes:
\begin{multline}
\label{amp}
\mathcal{A}(\Gamma, \{\varepsilon(s)\} ) \,=\, \\N^{2N_0 + N_1 + N_2}
     \sum_{\{\hat{I}(\rho^{1})\} \in P_+^1(\hat{\mathfrak{g}})}
    \prod_{\{\rho^{0}(p,q,r)\}}^{N_{2}(T)} 
  \begin{Bmatrix}
    \hat{I}_1(q,p) & \hat{J}_1(p) & \hat{J}_2(q)  \\
    \hat{J}_3(q) & \hat{I}_2(r,s) & \hat{I}_3(q,r)
  \end{Bmatrix}_{Q\,=\,e^{\frac{2 \pi i}{1 + h^\vee}}} \\
    \prod_{\{\rho ^{1}(p,r)\}}^{N_{1}(T)}\left(
      b_{\hat{I}(r,p)}^{\hat{J}(p)\hat{J}(r)}\right) ^{2}L(p,r)^{-2H_{\hat{I}(r,p)}}
\prod_{s=1}^{N_0(T)}
\sum_{\hat{K} \in P_+^1(\hat{\mathfrak{g}})}
\mathcal{N}_{\hat{I}(s)\,\hat{I}(s)}^{\hat{K}}
\text{Tr}_{\mathcal{H}_{\hat{K}}}[\tau_{h(s)}q^{L_0^{(O)} - \frac{p+1}{24}}]
\\
\prod_{m=1}^{D-p-1} \left[
     \sum_{\{\hat{j}(\rho^{1})\} \in P_+^1(\hat{\mathfrak{su}}(2))}
    \prod_{\{\rho^{0}(p,q,r)\}}^{N_{2}(T)} 
  \begin{Bmatrix}
    j_1(q,p) & j_1(p) & j_2(q)  \\
    j_3(q) & j_2(r,s) & j_3(q,r)
  \end{Bmatrix}_{Q\,=\,e^{\frac{2 \pi i}{1 + h^\vee}}} \right.\\\left.
    \prod_{\{\rho ^{1}(p,r)\}}^{N_{1}(T)}\left(
      b_{j(r,p)}^{j(p)j(r)}\right) ^{2}L(p,r)^{-2H_{j(r,p)}}
\prod_{s=1}^{N_0(T)}
\sum_{j \in P_+^1(\hat{\mathfrak{su}}(2))}
\mathcal{N}_{j(s)\,j(s)}^j
\text{Tr}_{\mathcal{H}_j}[\tau_{h(s)}q^{L_0^{(O)} - \frac{1}{24}}]
  \right]_{(m)},
\end{multline}
where the factor $N^{2N_0 + N_1 + N_2}$ takes into account the
degeneracy provided by the kinematical $U(1)^N$ Chan-Paton degrees of
freedom. 

The above scenario provides hopeful perspectives for its
generalization to the non Abelian case, in which the $U(N)$ symmetry
imposed in section \ref{sec:osgauge} is not broken. To this end, a key
point will be to look for a new consistent (although equivalent to
\eqref{bkBG}) identification between the background matrix
components and the entries of the Cartan matrix of the affine algebra
underlying the WZW-model. In particular, we do expect that the
extension to the non Abelian case may alter the $N^{2N_0+N1+N2}$
degeneracy factor appearing in \eqref{amp}. From a broader
perspective, it would be interesting to look for a dictionary between
the geometrical parameters underline the full construction in this
manuscript and physical quantities of a full-fledged bosonic string
theory. We hope to address such issues in a future paper.

\begin{center}
\subsection*{Acknowledgments}
\end{center}
  V.L.G. would like to thank the Foundation Boncompagni-Ludovisi,
  n\'ee bildt, for financially supporting her stay at Queen Mary,
  University of London

\newpage

\appendix

\section{Useful formulae}

This section contains a collection of useful equations and formulae
which can be found in standard group and algebra theory textbooks
such as \cite{Gilmore,Fulton-Harris,Hamermesh}

\begin{itemize}
\item {\bf Direct products}\\
  If a group G is a direct product of groups $G = G_1 \times G_2$,
  then, given any two elements $g_1 \in G_1$ and $g_2 \in G_2$, then a
  representation $R$ of $G$ can be written as 
  \begin{equation}
\label{eq:dir_prod}
    R^{\hat{I}_1 \times \hat{I}_2}_{m_1 n_1; m_2 n_2} (g_1 g_2) \,=\,
    R^{(1)\hat{I}_1}_{m_1 n_1}(g_1)\,
    R^{(2)\hat{I}_2}_{m_2 n_2}(g_2),
  \end{equation}
being $R^{(1)}$ and $R^{(2)}$ a representation respectively of $G_1$ and
$G_2$.
\item {\bf Clebsh-Gordan expansion}\\
Let us consider the expansion of the Kronecker product of two
representations:
\begin{equation*}
  R^{\hat{I}_1} \, R^{\hat{I}_2} \,=\, \sum_{\hat{I}\in
  P^+_k(\mathfrak{g})}
  (\hat{I}_1\,\hat{I}_2\,\hat{I}) R^{\hat{I}},
\end{equation*}
where $(\hat{I}_1\,\hat{I}_2\,\hat{I})$ is the number of times that
$R^{\hat{I}}$ enters in the Kronecker product of $R^{\hat{I}_1}$ and
$R^{\hat{I}_2}$.

Now let us consider the product of two representation functions with
the same argument. It can be expanded in the Clebsh-Gordan series:
  \begin{multline}
\label{eq:CG_series}
    R^{\hat{I}_1}_{m_1\,n_1}(\Gamma) \,  
    R^{\hat{I}_2}_{m_2\,n_2}(\Gamma) \,=\, \\
    \sum_{\hat{I} / (\hat{I}_1\,\hat{I}_2\,\hat{I}) \neq 0} 
    \sum_{M, N = 1, \ldots, \text{dim}|\hat{I}|}
    C^{\hat{I}\,M}_{\hat{I}_1\, m_1\, \hat{I}_2\, m_2\,}\,
    D^{\hat{I}}_{M\,N}(\Gamma)\,
    C^{\hat{I}\,N}_{\hat{I}_1\, n_1\, \hat{I}_2\, n_2\,},\,
  \end{multline}
where the sum is extended to those unitary representations for which
the coefficient $(\hat{I}_1\,\hat{I}_2\,\hat{I})$ is non zero.
\item {\bf Completeness relations for Clebsh-Gordan coefficients}\\
  \begin{subequations}
    \label{eq:CG_unit}
    \begin{gather}
      \label{eq:CG_unit1}
      \sum_{m_1\,m_2}
      C^{\hat{I}\,m}_{\hat{I}_1\,m_1\,\hat{I}_2\,m_2}
      C^{\hat{I}'\,m'}_{\hat{I}_1\,m_1\,\hat{I}_2\,m_2}
      \,=\,
      \delta_{\hat{I}\,\hat{I}'}\delta_{m\,m'}, \\
      \label{eq:CG_unit2}
      \sum_{\hat{I}\,m}
      C^{\hat{I}\,m}_{\hat{I}_1\,m_1\,\hat{I}_2\,m_2}
      C^{\hat{I}\,m}_{\hat{I}_1\,m_1'\,\hat{I}_2\,m_2'}
      \,=\,
      \delta_{m_1\,m_1'}\delta_{m_2\,m_2'}.
    \end{gather}
  \end{subequations}
\end{itemize}

\end{document}